\documentclass[12pt]{article}
\usepackage[x11names,dvipsnames,table]{xcolor}
\usepackage{graphicx}
\usepackage{caption}
\usepackage{newtxtext}
\usepackage{newtxmath}
\usepackage{subcaption}
\usepackage{hyperref}
\usepackage{mathtools}
\hypersetup{
    colorlinks = true,
    urlcolor   = blue,
    citecolor  = black,
}
\usepackage{epstopdf, epsfig}
\usepackage{amsmath}
\usepackage{cleveref}
\Crefname{figure}{Fig.}{Figs.}
\usepackage{float}
\usepackage{pgfplots}
\usepackage{natbib}
\usepackage{tikz}
\usetikzlibrary{positioning}

\usetikzlibrary{patterns.meta}
\usetikzlibrary{decorations,decorations.markings,decorations.text}
\usetikzlibrary{decorations.pathmorphing,arrows,intersections}
\usetikzlibrary{shadings}
\usepgflibrary{shadings}
\usepackage{tikz-dimline}

\newcommand{\RomanNumeralCaps}[1]

\title{Dynamics of two Interacting Drops in a Microfluidic Confinement under
imposed Temperature Gradient}

\author{
  Sayak Ray\thanks{Corresponding author: \texttt{sayokroy56@gmail.com}}, 
  Sudipta Ray, 
  Suman Chakraborty \\
  \normalsize Department of Mechanical Engineering, \\
  \normalsize Indian Institute of Technology, Kharagpur, India
}
\begin{document}
\maketitle


\begin{abstract}
Thermocapillary motion is widespread in both natural and engineering applications. A tiny drop of one liquid, suspended within another, may be set into motion aligned with an imposed thermal gradient, as influenced by thermocapillary action stemming from the gradients in interfacial tension due to the local variations in temperature. In real-world situations, however, such drops do not remain in isolation, as they interact with their neighboring entities, including other drops in proximity as well as a nearby solid boundary, setting up a complex interplay between the confinement-mediated interactions and the three-dimensional nature of the droplet dynamics. In this study, we present numerical solutions for the migration dynamics of a tightly confined drop couple, incorporating deformable interfaces, film flow, and Marangoni effects in the presence of dynamically evolving thermocapillary stresses induced by an imposed uniform temperature gradient. Unlike prior investigations, our work highlights the influence of the confinement towards orchestrating non-trivial features of drop migration, as dictated by an intricate coupling of the thermal and flow fields amidst the interferences of the domain boundaries. The study reveals that hydrodynamic interactions resulting from a juxtaposition of these influences deform the drops in a unique manner as compared to the characteristics evidenced by previously reported studies, causing a distortion of the local thermal fields around them. This, in turn, leads to changes in the local thermocapillary stress, affecting the local shear gradient in a manner that alters the local flow field in accordance with ensuring the interfacial stress balance. The consequent alteration in the drop velocities is shown to govern their migration in a distinctive manner, presenting unique signatures as compared to more restrictive scenarios studied previously. These findings hold significance in designing thermocapillary-driven micro-confined systems for controlling drop trajectories under an imposed thermal field, bearing far-reaching implications in a plethora of overarching applications ranging from droplet microfluidics to space technology.
\end{abstract}
\section{Introduction}
Thermocapillary motion is pervasive in both natural phenomena and engineering applications. A small drop of one fluid, suspended in another medium having a prevailing temperature gradient, may exhibit movement in the direction of the imposed gradient, as dictated by a resulting thermocapillary action \citep{anderson1985}. This phenomenon is typically orchestrated by the fact that the presence of a local temperature gradient generates a corresponding gradient of the interfacial tension along the surface of the drop. This differential tension acts as a force, pulling the surrounding fluid and propelling the drop towards areas where its interfacial tension would typically be lower, often in the direction of the hotter regions. Such phenomena, which have been intriguing to physicists over the years, have become progressively more important over the past decades from their application-oriented perspectives due to unprecedented advancements in miniaturization and space technology where several utilities are to function in near-weightless conditions. For example, the removal of unwanted liquid drops in a continuous phase by thermocapillary forces may greatly facilitate the processing of materials in outer space, minimizing various defects that are otherwise inevitable due to gravity-induced fluid phase segregation  \citep{uhlmann1981,carruthers1983,ostrach1982}. It is not far beyond imagination that the cooling system of space habitats may be achievable using thermocapillary migration. With the advent of microfluidics and miniaturization, thermocapillary phenomena have been attracting attention in several on-earth applications as well, such as micro heat pipes for thermal management of electronic equipment (Van Erp et al. 2020; Tang et al. 2018), where gravity-induced effects render to be inconsequential due to their large surface area by volume ratios. In several such scenarios, uncontrolled accumulation of drops via thermocapillary motion may be rather undesirable, as they may deteriorate the heat exchange efficacy between the hot and the cool interfaces. Imposing a delicate control over the thermocapillary migration of drops, therefore, appears to be imperative, irrespective of whether their motion needs to be accelerated or retarded.

A vast body of reported research on thermocapillary motion concerns the migration of single drops or bubbles in isolation \cite[]{young1959,hetsroni1970,Balasubramaniam1987,haj1997thermocapillary,chan1979,haj1990,zhang2001,wu2012}. For an accounting of the early studies in this field, one may refer to the review papers by \citet{subramanian2002} and \citet{wozniak1988}. While early investigations on this topic considered drops in unbounded flows, subsequent endeavors probed more closely the effects of the confining walls \cite[]{brady2011} on the drop dynamics. A noteworthy finding regarding the impact of boundary effects on thermocapillary motion was that a drop with significant thermal conductivity can undergo faster migration near a free fluid surface compared to when it is isolated. Nevertheless, in real-world applications, managing numerous bubbles or drops is often essential, and their collective behavior may deviate significantly from the intuitive expectations based on the individual particle outcomes, so understanding the dynamics of interacting drops renders it practically more imperative. 
In several practical scenarios, a drop interacts simultaneously with the domain boundaries and other neighboring drops \cite[]{keh1992}. These interactions may introduce strong local variations in the temperature gradients on the interfaces of the drops, leading to localized changes in the surface tension. The consequent alterations in the interfacial surface tension gradient-driven fluid motion near the interfaces may perturb the drop’s shape evolution and motion simultaneously as the interfaces are drawn in the direction of increasing interfacial tension. The consequent interaction of the drops, in lieu, may perpetually modify the local interfacial interactions (stress jump conditions) in a manner that may cause significant drop deformations even in the case of negligible convective transport. Further to this end, when attracted in sufficiently close vicinity by virtue of local gradients in interfacial tension, the interacting drops may coalesce as well, as observed in different natural and industrial processes, including liquid-liquid phase separation, polymer casting, and the treatment of liquid phase-miscibility-gap materials. For situations in which such coalescence renders undesirable such as the ones considered in this work, a careful a-priori rationalization of the drop interaction dynamics renders critical, in line with the intended drop migration features consistent with the particular application on focus.

In the literature, initial studies on the interaction of drops in the course of their thermocapillary migration were performed under the assumption of negligible deformation (capillary number tends to zero in the limit as applicable for quiescent flows of large surface tension or low viscosity fluids); for example, see the articles of \citet{meyyappan1984,meyyappan1983} and \citet{acrivos1990}. The motion of two liquid drops oriented arbitrarily with respect to a temperature gradient was examined analytically by \citet{anderson1985} in the low Reynolds and Marangoni number limit.  By using the two-drop solution, he also showed that the mean velocity of a drop suspension is lower than for a single drop. \citet{keh1990} examined the axisymmetric thermocapillary motion of two spherical drops progressing along their line of centers within a creeping flow. Their findings demonstrated that two identical liquid drops exhibit a faster migration compared to a single drop of the same size. Conversely, in the case of two gas bubbles with equal radii, no interaction was observed for all separation distances, aligning with the predictions made by \citet{meyyappan1984}. Later \citet{keh1992} explored the axisymmetric migration of a series of spherical drops and gas bubbles moving along their line of centers. In the case of multiple gas bubbles, it was demonstrated that the migration velocity of each bubble remained unaffected by the presence of the other bubbles if the bubbles were of the same size. \citet{wei1993} investigated theoretically the quasi-static thermocapillary migration of a chain of two and three spherical bubbles for zero Marangoni and Reynolds numbers. \citet{keh1992} considered the migration of drops oriented arbitrarily with respect to the temperature gradient in the limit of zero Marangoni and Reynolds numbers. Unlike drops moving along their line of centers \cite[]{keh1990} drops moving with their line of centers orthogonal to the temperature gradient were shown to migrate slower than a single drop. \citet{loewenberg1993} studied the axisymmetric, thermocapillary-driven motion of a pair of non-conducting, spherical drops in near-contact for small Reynolds and Marangoni numbers. Their study involved computing the pairwise motion and associated contact forces by examining touching drops in point contact. In this scenario, the relative motion between nearly touching drops initiated from the contact force, which was counteracted by a lubrication resistance. The conclusion drawn was that, for nearly equi-sized drops, the ratio of relative velocity between two drops in near contact to that for widely separated drops is consistent for both thermocapillary-driven and gravity-driven motion.

The interaction of two deformable drops in the axisymmetric coordinates was studied by \citet{zhou1996}. Numerical simulations of an axisymmetric buoyancy-driven interaction of a leading drop and a smaller trailing drop were reported by \citet{zinchenko1999}. This study revealed that the trailing drop experiences significant elongation due to the hydrodynamic influence exerted by the leading drop. Subsequently, depending on the governing parameters, the drops may either separate and revert to a spherical shape, the trailing drop may be captured by the leading one, or one of the drops may undergo breakup. In the context of thermocapillary-induced motion, the impact of deformability was primarily investigated using a perturbation technique, assuming small deformations. \citet{rother1999} applied lubrication approximation to study the effect of slight deformability of the interfaces on the thermocapillary-driven migration of two drops at close proximity.

The investigations of interactions of drops discussed above have all been limited to zero Reynolds and Marangoni numbers. \citet{nas2003} conducted a computational investigation into the thermocapillary migration of two fully deformable bubbles and drops, considering non-zero values of the Reynolds and Marangoni numbers. The results indicated that the bubbles and lightweight drops got aligned perpendicularly to the temperature gradient and were uniformly spaced in the horizontal direction. A space experiment evidenced that a small leading drop could retard the movement of the big trailing drop in the process \cite[]{balasubramaniam1996}. \citet{yin2011} studied the thermocapillary interaction of two arbitrarily placed drops, considering them to be of the same size and having the same physical parameters (kinematic viscosity, thermal diffusivity, density, and specific heat).

One critical feature that was shown to demarcate the characteristics of two interacting drops as compared to the corresponding single-drop dynamics is the distinction between their impending coalescing regime and in-tact motion. Of great interest is the non-coalescing behavior of the interacting drops, which had its early foundation in the seminal studies of Lord Rayleigh on the behavior of water jets that bounce over one another \cite[]{rayleigh1899}, with its resurgence about a century later in the form of surface vibration-facilitated non-coalescence \cite[]{walker1978} that fundamentally aimed to inhibit the impending drainage of liquid between the two interacting drops in close proximity  \cite[]{marrucci1969,anilkumar1991}. Systems exhibiting this apparently unusual non-coalescence thence continued to attract attention, particularly for their implications in materials science, meteorology, and microgravity experiments \cite[]{fredriksson1984}. Under thermocapillary effects, the enhanced interfacial shear due to Marangoni effects, opposing the draining of the film between two interfaces, may resist the interfacial tension that facilitates coalescence, resulting in a dynamic enhancement of the resistance to drainage. Therefore, thermocapillary effects assume significance among the possible means that may be harnessed for the controlled movement of multiple interacting drops without having them intermingled \cite{dell1996}. However, addressing such problems from a theoretical perspective remains challenging. The challenges are attributable to a number of factors: the non-negligible drop deformation, the three-dimensionality of the transport stemming from 
confinement-induced interactions, and the dynamics of the thermocapillary stress field due to the spontaneously varying temperatures around the interacting drops. Most imperatively, these key factors do not act in isolation but have a non-trivial interplay because of a two-way coupling between the heat transfer and fluid flow as mediated by an interfacial stress balance that delves into a dynamically evolving thermal field around the interacting drops. Whereas the analytical techniques put forward to address the dynamics of a droplet couple promised to be richly insightful to an extent, they appeared to be clearly inadequate in deciphering the resulting complex coupling, establishing the need for more exhaustive computational frameworks.

For moderate and large drop deformations, the boundary-integral method, as pioneered by \citet{rallison1978} and described in detail by \citet{pozrikidis1992}, emerged to be of utilitarian importance for analyzing the hydrodynamic problem in the Stokes flow limit. \citet{zhou1996} conducted a study on the asymmetric motion of two deformable drops subjected to a temperature gradient along the line of their centers. Disregarding heat convection and inertial effects, they computed the temperature and velocity fields for significant drop deformations using boundary-integral techniques for the Laplace and the Stokes equations, respectively. They presented detailed numerical results on drop motion, deformation, and the temporal evolution of the gap width between the drops, considering equal viscosities of the drops and surrounding fluid. The study highlighted the influence of the capillary number, drop size ratio, and drop-to-medium conductivity ratio on drop motion and deformation. Their results indicated that hydrodynamic interactions between the drops exerted a more pronounced effect on the smaller of the two drops, impacting both drop motion and deformation. Deformation was shown to significantly affect the rate of thin film drainage between the drops, while its impact on the velocities of the drop centers was relatively marginal. To the limit of their calculations, they were able to confirm the predictions of \citet{loewenberg1993}. \citet{berejnov2001} analyzed the problem with the trailing drop smaller than or equal to the leading drop. \citet{zhou1996} used a boundary-integral technique to study the thermocapillary interaction of a deformable viscous drop with a larger trailing drop making no a priori assumptions regarding the magnitude of deformations, inferring that the influence of deformability became significant only when the drops came close to each other. However, they did not consider the variations in the surface tension due to continuous changes in the positions of the drops. \citet{berejnov2001} investigated the influence of the deformations on the relative motion of the drops in the case of moderate capillary numbers and equal viscosity and thermal properties of the dispersed and continuous phases.  \citet{rother2002} generalized the above axisymmetric analyses to three dimensions and arbitrary viscosity ratio by adapting the boundary-integral code of \citet{zinchenko1999} to handle the tangential Marangoni stresses. \citet{lavrenteva2001} probed the scenarios of high Peclet numbers for analyzing the thermocapillary interaction among drops. However, the confinement effects on two-drop thermocapillary interaction amidst a dynamically evolving Marangoni stress acting on them remained to be addressed thus far.

Here we arrive at three-dimensional numerical solutions of the Stokes equation, with deformable interfaces, film flow, and the Marangoni effects in the presence of dynamically evolving thermocapillary stresses on the application of a uniform temperature gradient on a micro-confined drop pair. In contrast to previous investigations, our work puts forward the effects of confinement amidst the coupled thermal and flow fields in a boundary element framework. The hydrodynamic interactions due to the confining boundaries are shown to deform the drops from their respective equilibrium shapes, which results in a distortion of the local thermal field around their neighbourhoods. This, in turn, alters the local thermocapillary stress and, consequently, the local shear gradient to ensure the interfacial stress balance. The resulting alteration in the flow field is shown to dictate the migration of the drops in an intriguing manner having distinctive signatures compared to other more restrictive scenarios studied previously. These results are likely imperative in designing thermocapillary-driven micro-confined systems for controlled drop trajectories under an imposed thermal field. 

\section{Problem Formulation}
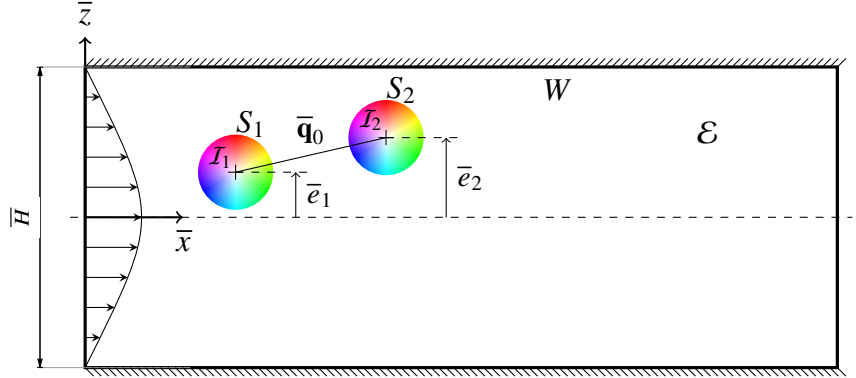
\begin{figure}
\begin{center}
            \begin{tikzpicture}[xscale = 2, yscale =2, plotmark/.style = {%
           draw, fill=red, circle, inner sep=0pt, minimum size=4pt}]
        \draw[black,very thick] (0,0) rectangle (5,2);
        \shade[shading=color wheel white center] (2,1.53) circle (0.25);
        \draw (1.95,1.53) -- (2.05,1.53);
        \draw (2,1.48) -- (2,1.58);
        \shade[shading=color wheel white center] (1,1.3) circle (0.25);
        \draw (0.95,1.3) -- (1.05,1.3);
        \draw (1,1.25) -- (1,1.35);
        \draw[name path=curve] (0,0) .. controls (0.5,1) .. (0,2);
        \foreach \y in {0.2,0.4,...,1.9}{
          \path[name path=horizontal] (0,\y) -- + (3,0);
          \draw[-stealth,name intersections={of=curve and horizontal}] (0,\y) -- (intersection-1);}
        \draw[->,thick] (0.0,1) -- (0.65,1) ;
        \draw[->,thick] (0.0,1) -- (0.0,2.2)node[above] {$\overline{z}$};
        \draw (0.66, 0.85)  node {$\overline{x}$};
        \draw (4.14,1.56) node {$\mathcal{E}$};
        \draw (1.1,1.62) node {$S_1$};
        \draw (2.1,1.85) node {$S_2$};
        \draw (1.9,1.65) node[scale=0.8] { $\mathcal{I}_2$};
        \draw (0.9,1.4) node [scale=0.8] { 
        $\mathcal{I}_1$};
        \draw (3.14,1.85) node {$W$};
        \coordinate (A) at (-0.3,0);
        \coordinate (B) at (3,0);
        \coordinate (C) at (3,2);
        \coordinate (D) at (-0.3,2);
        \coordinate (E) at (4.2,0);
        \coordinate (F) at (4.2,2);
        \draw (A)--(D);
        \coordinate (A1) at (2,1.3);
        \coordinate (B1) at (1.4,1.3);
        \coordinate (D1) at (1.4,1);
        \coordinate (A2) at (2.4,1);
        \coordinate (B2) at (2.4,1.53);
        \coordinate (ce1) at (1,1.3);
        \coordinate (ce2) at (2,1.53);
         \dimline[label style ={above=0.00004ex,font=\scriptsize}]{(A)}{(D)}{$\overline{H}$};
         \draw[dashed] (1,1.3) -- (1.5,1.3);
         \draw[dashed] (2,1.53) -- (2.5,1.53);
         \draw[->](D1)--(B1) node [midway,anchor=west]{\small $\overline{e}_1$};
         \draw[->](A2)--(B2) node [midway,anchor=west]{\small $\overline{e}_2$};
         \draw (ce1)--(ce2) node [midway,anchor=south]{\small $\overline{\mathbf{q}}_0$};
        \draw[dashed] (-0.1,1) -- (5.1,1);
        \foreach \x in {0.0,0.05,...,5.05}{
          \draw[rotate around={-45:(\x,2)}] (\x,2) -- (\x,2.08);}    
        \foreach \x in {0.0,0.05,...,5.05}{
          \draw[rotate around={-135:(\x,0)}] (\x,0) -- (\x,0.08);}   
        \end{tikzpicture}   .
        \end{center}
\caption{Schematic diagram of the confined channel with two droplets in an aligned arrangement}
\label{setup}
\end{figure}

We consider two Newtonian droplets of density $\overline{\rho}_i$, thermal conductivity $\overline{k}_i$, specific heat $\overline{c}_{i}$, viscosity $\overline{\mu}_i$ suspended in a fluid of density $\overline{\rho}_e$, thermal conductivity $\overline{k}_e$, specific heat $\overline{c}_{e}$, viscosity $\overline{\mu}_e$. The domain of the imposed flow is a parallelepiped channel $\mathcal{E}$ (ref. \Cref{setup}), confined in the $z$-direction by wall $W$ and the spherical droplets $\mathcal{I}_1$ and $\mathcal{I}_2$ are suspended in the channel. The channel length in the $x$ and $y$-direction is much larger than in the $z$-direction. The imposed flow is assumed to be a fully developed Poiseuille flow. The initial droplet radius is $\overline{a}$, and the surface of each droplet is denoted as $S_l, \, l = 1,2$. The instantaneous position of the centroid of the $l^{th}$-droplet is given by  $\overline{\mathbf{x}}_{c,l}(\overline{t}) = \{\overline{x}_{c,l},\,\overline{y}_{c,l}, \,\overline{z}_{c,l}\}$, whereas, the initial position of the droplet is $\overline{\mathbf{x}}^{0}_{c,l} = \{\overline{x}^{0}_{c,l},\,\overline{y}^{0}_{c,l}, \,\overline{z}^{0}_{c,l}\}$. The droplets are initially separated by $\mathbf{\overline{q}}_{0} = \{\overline{d}_0,\overline{g}_0,\overline{h}_0\}$ where $\overline{d}_0 = \overline{x}^{0}_{c2}-\overline{x}^{0}_{c1}$, $\overline{g}_0 = \overline{y}^{0}_{c2}-\overline{y}^{0}_{c1}$ and $\overline{h}_0 = \overline{z}^{0}_{c2}-\overline{z}^{0}_{c1}$. The initial $z$-offset of the droplets from the $x$-axis is given as $\overline{e}_{l}$ and the initial $y$-offset is given as $\overline{p}_{l}$. The velocity of the droplet centroid is  $\overline{\mathbf{u}}_l = \{\overline{U}_{x,l},\,\overline{U}_{y,l},\,\overline{U}_{z,l}\}$ and the temperature field is given by $\overline{T}(\mathbf{x})$. We define $\overline{T}_{sl} = \overline{T}(\mathbf{x}) \text{ for } x \in S_l,\, l = 1,2$. The 
Throughout the discussion, the suffix $e$ will denote the bulk fluid, whereas $i$ will denote the values corresponding to the droplets, and we have used the suffix $s$ to indicate the surface of the droplets.
A temperature gradient of $d\overline{T}/d\overline{x} = \overline{G}_{a}$ is considered to be present in the continuous phase along the flow. The surface tension $\overline{\sigma}$ decreases linearly with droplet surface temperature $\overline{T}_{s}$ \citep{young1959}, from a reference value of $\overline{\sigma}_{0}$ at a reference temperature $\overline{T}_{0}$. The slope of the surface tension with temperature is given as $\beta$. 
\subsection{Normalization}
Normalization of the space is essential to understand the functional parameters affecting the problem. For the present physical problem, the length scale is the channel height ($\overline{H}$), whereas the velocity scale is the magnitude of the centreline velocity $\overline{\mathbf{v}}_c = \overline{V}_c\mathbf{\hat{x}}$ of the external flow, where $\mathbf{\hat{x}}$ is the unit vector in the x-direction. Therefore, we can implicitly define a time scale for the present problem as $\overline{H}/\,\overline{V}_c$. For the thermal problem, the temperature scale is taken to be $|\overline{G}_{a}|\overline{H}$ as the scaled temperature difference $\Delta \overline{T}$. The non-dimensional quantities are given as 
\begin{align*}
    x &= \frac{\overline{x}}{\overline{H}} &
    y &= \frac{\overline{y}}{\overline{H}} &
    z &= \frac{\overline{z}}{\overline{H}} & 
    T &= \frac{\overline{T}-\overline{T_{0}}}{|\overline{G_{a}}|\overline{H}} \\
    u_x &= \frac{\overline{u}_x}{\overline{V}_c} &
    u_y &= \frac{\overline{u}_y}{\overline{V}_c} &
    u_z &= \frac{\overline{u}_z}{\overline{V}_c} & 
    \sigma &= \frac{\overline{\sigma}}{\overline{\mu}_{e}\overline{V}_c}
\end{align*} 
For the thermocapillary-induced migration problem, the following non-dimensional parameters are of particular importance:
\begin{enumerate}
    \item Capillary number given as $Ca = \overline{\mu_e{V}}_c/\,\overline{\sigma}_0$. This parameter denotes the relative importance of the viscous and surface tension forces. The deformation of the droplets is small for the small values of the Capillary number considered here. 
    \item \emph{Marangoni number} given as $Ma = \beta|\overline{G}_a|\overline{H}/\,\overline{\mu_eV}_c$. This parameter denotes the relative importance of the temperature-driven Marangoni flow over the strength of the imposed flow. This number is also kept small for the particular problem. 
    \item Thermal \emph{Peclet number} given as $Pe = \overline{V_cH}/\,\overline{\alpha}_e$. this gives the relative importance of the convective transport over the diffusive transport of heat. This is considered negligible to signify that the motion of the drops is entirely diffusive. This is observed in several systems of practical interest \citep{nallani1993}.
    \item \emph{Reynolds number} is given as $Re = \overline{\rho_eV_cH}/\,\overline{\mu}_e$. This shows the relative importance of the convective transport over the diffusive transport of momentum. The Reynolds number is assumed to be negligible, and hence a highly viscous flow is considered, which is commonly referred to as the "creeping-flow approximation".     
    \item Confinement ratio ($Cr$), a measure of the droplet confinement within the channel, given as $Cr =2\overline{a}/\overline{H}$. For $Cr \sim 0.2$, the flow can be assumed to be unbounded \citep{keh2002}. 
    \item Initial droplet separation ($\mathbf{q}_{0}$), given as $\mathbf{q}_{0} = \overline{\mathbf{q}}_{0}/\overline{H}$.
    \item Droplet offset distances are the initial offsets for the $l^{th}$ droplet, given as $e_{l} = \overline{e}_{l}/\overline{H}$, $p_{l} = \overline{p}_{l}/\overline{H}$. 
    
\end{enumerate}
\subsection{Governing Equations and boundary equations}

The non-dimensional momentum equation for the outside fluid is given as 
\begin{equation}\label{vel_eqn_ext}
    Re\left(\frac{\partial \mathbf{u}}{\partial t}+(\mathbf{u}.\nabla)\mathbf{u}\right) = -\nabla P +\nabla^2\mathbf{u}, \, \forall \mathbf{x} \in \mathcal{E} 
\end{equation}
while for the interior of the droplets $\mathcal{I} \equiv \mathcal{I}_1 \cup \mathcal{I}_2$, the velocity distribution is obtained from
\begin{equation}\label{vel_eqn_int}
    Re\rho_r\left(\frac{\partial \mathbf{u}}{\partial t}+(\mathbf{u}.\nabla)\mathbf{u}\right) = -\nabla P +\mu_r\nabla^2\mathbf{u}, \,  \forall \mathbf{x} \in \mathcal{I} 
\end{equation} 
Here, $\rho_r = \overline{\rho}_i/\,\overline{\rho}_e$ and $\mu_r = \overline{\mu}_i/\,\overline{\mu}_e$. For the present work, $\rho_r = \mu_r = 1$. The dimensionless time is $t \equiv \overline{t\mathbf{v}}_c/\,\overline{H}$. 

The temperature distribution in the outside fluid is obtained by solving the thermal energy equation
\begin{equation}\label{heat_eqn_ext}
   Pe \left(\frac{\partial T}{\partial t}+(\mathbf{u}. \nabla)T\right) = \nabla^2 T, \,  \forall \mathbf{x} \in \mathcal{E}
\end{equation}
whereas for the droplet interior, we solve the following 
\begin{equation}\label{heat_eqn_int}
    \rho_{r}c_{r}Pe\left(\frac{\partial T}{\partial t}+(\mathbf{u}.\nabla)T\right) = \delta\nabla^2 T,  \,  \forall \mathbf{x} \in \mathcal{I}
\end{equation}
Here $c_{r} = \overline{c}_i/\overline{c}_e$.
\par The Reynolds numbers and Peclet numbers are assumed to be negligible for the present work, and assuming a steady state, we simplify the energy equation (\Cref{heat_eqn_ext}) for the external fluid as 
\begin{equation}\label{heat_eq_1}
  \nabla^2 T = 0, \,  \forall \mathbf{x} \in \mathcal{E}
\end{equation}
and for the droplets (\Cref{heat_eqn_int}), we obtain  
\begin{equation}\label{heat_eq_2}
  \nabla^2 T = 0 
 \,\,  \forall \mathbf{x} \in \mathcal{I}
\end{equation}
The thermal boundary conditions at the interface of the droplets and main fluid are  
\begin{equation}\label{heat_bd_1}
\begin{rcases} T_{i} = T_{e} = T_{s}, \\  \delta(\mathbf{n}.\nabla T_{i}) = \mathbf{n}.\nabla T_{e}, \end{rcases}\, \mathbf{x} \in S_l
\end{equation} and 
$T_s$ is the non-dimensional surface temperature
$\mathbf{n}$ denotes the normal unit vector at the $l^{th}$-interface. 
Here, $\delta = \overline{k}_i/\overline{k}_e$ is the conductivity ratio between the fluid inside the droplet and the continuous medium. The imposed temperature field ($T_\infty$) is given as 
\begin{equation}\label{heat_bd_2}
    T_{\infty}(\mathbf{x}) = \gamma x,
\end{equation}
where $\gamma$ can be +1 or -1 depending on the direction of the temperature field. 
The walls ($W$) are considered to be insulated
\begin{equation}\label{heat_bd_3}
    \mathbf{n}.\nabla T = 0, \, \mathbf{x} \in W.
\end{equation} 
For the present study, given the steady-state nature of the flow as well as the vanishingly small Reynolds number considered here, the flow governing equation (\Cref{vel_eqn_ext}) simplifies to the \emph{Stokes flow} equation for the external flow, 
\begin{equation}\label{vel_eqn_1}
    \nabla P = \nabla^2 \mathbf{u},\,\,  \forall \mathbf{x} \in \mathcal{E}
\end{equation}
and for the fluid medium inside the droplets, considering $\mu_{r} = 1$, \Cref{vel_eqn_int} reduces to
\begin{equation}\label{vel_eqn_2}
    \nabla P = \nabla^2 \mathbf{u},\,\,  \forall \mathbf{x} \in \mathcal{I}
\end{equation}
The continuity equations are given as
\begin{equation}\label{vel_eqn_3}
    \nabla. \mathbf{u} = 0, \,\,  \forall \mathbf{x} \in \mathcal{E}
\end{equation}
\begin{equation}\label{vel_eqn_4}
    \nabla. \mathbf{u} = 0, \,\,  \forall \mathbf{x} \in \mathcal{I}
\end{equation}
The imposed flow is considered to have a fully-developed profile,
\begin{equation}\label{vel_inf}
\mathbf{u}_\infty = (1-4z^2)\mathbf{\hat{x}}
\end{equation}
where $\hat{x}$ is the unit vector along the x-direction.
The solid walls at $\overline{z} = \pm \overline{H}/2$($z = \pm 0.5$) is no slip which means 
that at the walls
\begin{equation}\label{fluid_bd_1}
 \mathbf{u} = \mathbf{u}_{w} = 0, \, \mathbf{x} \in W   
\end{equation}
Here the boundary conditions at the interface between the external fluid and the droplets are given as 
\begin{equation}\label{fluid_bd_2}
\begin{rcases} \mathbf{u}_{i}.\mathbf{n} = \mathbf{u}_{e}.\mathbf{n} = \mathbf{u}_{s}.\mathbf{n} = \frac{d\mathbf{x}}{dt}.\mathbf{n}, \\  \mathbf{u}_{i} - (\mathbf{u}_{i}.\mathbf{n}).\mathbf{n}=\mathbf{u}_{e} - (\mathbf{u}_{e}.\mathbf{n}).\mathbf{n} \end{rcases} \, \mathbf{x} \in S_l \text{ for } \, l=1,2
\end{equation}
The traction equations at the interface of the fluid and the droplet. 
\begin{equation}\label{fluid_bd_3}
(\mathbf{S}_e-\mathbf{S}_{i}).\mathbf{n} = \sigma\mathbf{n}(\nabla.\mathbf{n})-\nabla_S\sigma, \mathbf{x} \, \in S_l \text{ for } \, l=1,2
\end{equation}
Here
\begin{equation}\label{nablaS}
\nabla_S = (\mathbf{I} - \mathbf{n}\mathbf{n}).\nabla
\end{equation}
The terms $\mathbf{S}_e$, $\mathbf{S}_{i}$ represent the stress tensors within the external and the droplets, respectively. The term $\nabla_S$ represents the projection of the gradient operator on the interface $S_{l}$. 
The surface tension is given by 
\begin{equation}\label{fluid_bd_4}
    \sigma (\mathbf{x}) = \frac{1}{Ca} - MaT_s
\end{equation}
Here $Ca$ is the capillary number of the droplets while the term $Ma$ is the Marangoni number as defined earlier. 
Using \Cref{fluid_bd_4} in \Cref{fluid_bd_3} along with \Cref{nablaS}, we get the following equation. 
\begin{equation}\label{fluid_bd_5}
(\mathbf{S}_{e}-\mathbf{S}_{i}).\mathbf{n}= Ma \nabla_S T_{s}+\left(\frac{1}{Ca}-Ma T_{s}\right)\mathbf{n}(\nabla.\mathbf{n}), \mathbf{x} \, \in S_l \text{ for } \, l=1,2
\end{equation}
The evolution of the droplet surface is given \Cref{fluid_bd_2}. 
The above equation can be written ignoring any phase change phenomenon happening at the interface of the droplets. The next section presents the discretisation of the governing equations and the implementation of the numerical method. 
\section{Numerical approach}
\subsection{Method overview}
The \emph{Boundary Element Method} (BEM) \citep{pozrikidis1992,pozrikidis2002,yon1998} is the numerical technique used here to solve the problem of thermocapillary migration of the two droplets in a confined domain. The 3D governing equations are reduced to surface integrals over the droplet interfaces, thereby potentially decreasing the overall computational cost. To solve the surface integrals, we generate a grid on the surface to compute the integrals instead of a volumetric grid. Since the focus of the present work is to observe the relative impact of the different driving forces, the BEM provides the ideal framework for the numerical study. We have further computed the values within the flow to visualize the overall flow pattern for a few cases using the surface evaluation of the field variables. 
\subsection{Discretization of Energy equation}
The \Cref{heat_eq_1,heat_eq_2} along with the boundary conditions \Cref{heat_bd_1,heat_bd_2,heat_bd_3} are discretized by the boundary element method.  The boundary integral equations formed from the equations are given below. For the sake of brevity, the full derivation of the boundary integral forms has been omitted. We define ${T}_{sl} = {T}(\mathbf{x}) \text{ for } x \in S_l,\, l = 1,2$ and $T_w$ denotes the temperature at the wall ($W$). The term $G(\mathbf{x},\mathbf{x_{0}})$ is the free space green's function for the Laplace operator and is given as
\begin{equation*}
G(\mathbf{x},\mathbf{x_{0}}) = \frac{1}{4\pi(|\mathbf{x} - \mathbf{x_0}|)}
\end{equation*}
The collocation point coordinate vector and the 
equation is given for the collocation point on the surface of the first droplet 
\begin{multline}\label{temp_1_ref}
\frac{T_{s1}(\mathbf{x}_{0})(1+\delta)}{2} = T_{\infty}+ (1-\delta)\int_{S_1}^{PV}(\mathbf{n}.\nabla G(\mathbf{x},\mathbf{x}_{0}))T_{s1} \,ds\\+(1-\delta)\int_{S_2}(\mathbf{n}\cdot\nabla G(\mathbf{x},\mathbf{x}_0))T_{s2} \,ds+\int_W (\mathbf{n}.\nabla G(\mathbf{x},\mathbf{x}_0))T_w\,ds+\int_W (\mathbf{n}\cdot\nabla T)G(\mathbf{x},\mathbf{x}_{0})\,ds
\end{multline}
For the collocation point on the second droplet surface 
\begin{multline}\label{temp_2_ref}
\frac{T_{s2}(\mathbf{x}_0)(1+\delta)}{2} = T_{\infty}+ (1-\delta)\int_{S_2}^{PV}(\mathbf{n}\cdot\nabla G(\mathbf{x},\mathbf{x}_0))T_{s2} \,ds\\+(1-\delta)\int_{S_1}(\mathbf{n}\cdot\nabla G(\mathbf{x},\mathbf{x}_0))T_{s1} \,ds+\int_W (\mathbf{n}\cdot\nabla G(\mathbf{x},\mathbf{x}_0))T_w\,ds+\int_W (\mathbf{n}\cdot\nabla T)G(\mathbf{x},\mathbf{x}_{0}).\,ds
\end{multline}
For the collocation point on the surface of the wall
\begin{multline}\label{temp_3_ref}
\frac{T_{w}(\mathbf{x}_0)}{2} = T_{\infty}+ (1-\delta)\int_{S_2}(\mathbf{n}\cdot\nabla G(\mathbf{x},\mathbf{x}_0))T_{s2} \,ds\\+(1-\delta)\int_{S_1}(\mathbf{n}\cdot\nabla G(\mathbf{x},\mathbf{x}_0))T_{s1} \,ds+\int_W^{PV} (\mathbf{n}\cdot\nabla G(\mathbf{x},\mathbf{x}_0))T_w\,ds+\int_W (\mathbf{n}\cdot\nabla T)G(\mathbf{x},\mathbf{x}_{0})\,ds
\end{multline}
These equations \Cref{temp_1_ref,temp_2_ref,temp_3_ref} can be used to solve for the temperatures ($T_{s1}$,$T_{s2}$ and $T_w$) after imposition of boundary conditions \Cref{heat_bd_2,heat_bd_3}. 
\subsection{Discretization of momentum equation}
For the velocity,  we have the integral formulation for flow across an interface for two liquids of equal viscosity. This formulation is for points on the interface of the droplet and the main fluid. Here $\mu = \overline{\mu}_e$ is the viscosity of the external fluid while $\mathbf{u}_{s1},\mathbf{u}_{s2}$ is the velocity over the $S_1$ and $S_2$ respectively. 
For the first droplet, we have the following velocity formulation.
\begin{multline}\label{vel_1}
\mathbf{u}_{s1}(\mathbf{x}_0) = \mathbf{u}_{\infty}(\mathbf{x}_0)-\frac{1}{8\pi\mu}\int^{PV}_{S_1}((\mathbf{S}_{e}-\mathbf{S}_{i})\cdot\mathbf{n})\mathbf{G}\,ds\\-\frac{1}{8\pi\mu}\int_{S_2}((\mathbf{S}_{e}-\mathbf{S}_{i})\cdot\mathbf{n})\mathbf{G}\,ds 
-\frac{1}{8\pi\mu}\int_{W}(\mathbf{S}_{w}\cdot\mathbf{n})\mathbf{G}\,ds
\end{multline}
For the second droplet, we have the following 
\begin{multline}\label{vel_2}
\mathbf{u}_{s2}(\mathbf{x}_0) = \mathbf{u}_{\infty}(\mathbf{x}_0)-\frac{1}{8\pi\mu}\int^{PV}_{S_2}((\mathbf{S}_{e}-\mathbf{S}_{i})\cdot\mathbf{n})\mathbf{G}\,ds\\-\frac{1}{8\pi\mu}\int_{S_1}((\mathbf{S}_{e}-\mathbf{S}_{i})\cdot\mathbf{n})\mathbf{G}\,ds 
-\frac{1}{8\pi\mu}\int_{W}(\mathbf{S}_{w}\cdot\mathbf{n})\mathbf{G}\,ds
\end{multline}
For the wall, we have 
\begin{multline}\label{vel_3_ref}
\mathbf{u}_{w}(\mathbf{x}_0) = \mathbf{u}_{\infty}(\mathbf{x}_0)-\frac{1}{8\pi\mu}\int_{S_2}((\mathbf{S}_{e}-\mathbf{S}_{i})\cdot\mathbf{n})\mathbf{G}\,ds\\-\frac{1}{8\pi\mu}\int_{S_1}((\mathbf{S}_{e}-\mathbf{S}_{i})\cdot\mathbf{n})\mathbf{G}\,ds 
-\frac{1}{8\pi\mu}\int^{PV}_{W}(\mathbf{S}_{w}\cdot\mathbf{n})\mathbf{G}\,ds
\end{multline}
Here the term $\mathbf{G}$ refers to the \emph{free-space Greens function} for the Stokes equation \citep{pozrikidis2002practical}. When the two droplets come very close to each other, the integral term corresponding to the surface of the other droplet (third term in the RHS of \Cref{vel_1,vel_2}) becomes singular. To alleviate this numerical issue, the near-singularity subtraction of the integrand $f(\mathbf{x})$ is performed \citep{zinchenko1997}
\begin{equation}
     \int_{S_l} f(\mathbf{x})\cdot\mathbf{n}(\mathbf{x})  \mathbf{G} \,ds = \int_{S_l} [f(\mathbf{x})-f(\mathbf{x^*})] \cdot\mathbf{n}(\mathbf{x}) \mathbf{G} \,ds
\end{equation}
where, $\mathbf{x^*}$ is the nearest collocation node to $\mathbf{x}_0$ on $S_l$ when $\mathbf{x}_0 \notin S_l$. These equations \Cref{vel_1,vel_2,vel_3_ref} can be used to solve for the velocity of points on the droplet after imposition of boundary conditions \Cref{vel_inf,fluid_bd_1,fluid_bd_2,fluid_bd_3}. 
\subsection{Imposition of boundary conditions}
The boundary conditions for the heat transfer equations, which are steady state heat conduction equations, are given as \Cref{heat_bd_2,heat_bd_3}. These boundary conditions, when imposed in \Cref{temp_1_ref,temp_2_ref,temp_3_ref} are given for the collocation point on the surface of the first droplet 
\begin{multline}\label{temp_1}
\frac{T_{s1}(\mathbf{x}_{0})(1+\delta)}{2} = T_{\infty}+ (1-\delta)\int_{S_1}^{PV}(\mathbf{n}\cdot\nabla G(\mathbf{x},\mathbf{x}_{0}))T_{s1} \,ds\\+(1-\delta)\int_{S_2}(\mathbf{n}\cdot\nabla G(\mathbf{x},\mathbf{x}_0))T_{s2} \,ds+\int_W (\mathbf{n}.\nabla G(\mathbf{x},\mathbf{x}_0))T_w\,ds
\end{multline}
For the collocation point on the second droplet surface 
\begin{multline}\label{temp_2}
\frac{T_{s2}(\mathbf{x}_0)(1+\delta)}{2} = T_{\infty}+ (1-\delta)\int_{S_2}^{PV}(\mathbf{n}\cdot\nabla G(\mathbf{x},\mathbf{x}_0))T_{s2} \,ds\\+(1-\delta)\int_{S_1}(\mathbf{n}.\nabla G(\mathbf{x},\mathbf{x}_0))T_{s1} \,ds+\int_W (\mathbf{n}.\nabla G(\mathbf{x},\mathbf{x}_0))T_w\,ds
\end{multline}
For the collocation point on the surface of the wall,
\begin{multline}\label{temp_3}
\frac{T_{w}(\mathbf{x}_0)}{2} = T_{\infty}+ (1-\delta)\int_{S_2}(\mathbf{n}\cdot\nabla G(\mathbf{x},\mathbf{x}_0))T_{s2} \,ds\\+(1-\delta)\int_{S_1}(\mathbf{n}\cdot\nabla G(\mathbf{x},\mathbf{x}_0))T_{s1} \,ds+\int_W^{PV} (\mathbf{n}\cdot\nabla G(\mathbf{x},\mathbf{x}_0))T_w \,ds
\end{multline}
These equations \Cref{temp_1,temp_2,temp_3} can be used to solve for the temperature on the surface of the two droplets and the wall surface. Now, we use the boundary conditions \Cref{fluid_bd_1} in \Cref{vel_1,vel_2,vel_3_ref}. To calculate the droplet's surface velocity, we have to use \Cref{vel_3_ref} for when the collocation point is at the wall to calculate the unknown traction term ($\mathbf{S}_{w}.\mathbf{n}$). From \Cref{fluid_bd_1}, the free stream velocity at the wall is zero ($\mathbf{u}_{\infty}$). This means that \Cref{vel_3_ref} reduces to 
\begin{multline}\label{vel_3}
0 = \mathbf{u}_{\infty}(\mathbf{x}_0)-\frac{1}{8\pi\mu}\int_{S_2}((\mathbf{S}_{e}-\mathbf{S}_{i})\cdot\mathbf{n})\mathbf{G}\,ds\\-\frac{1}{8\pi\mu}\int_{S_1}((\mathbf{S}_{e}-\mathbf{S}_{i})\cdot\mathbf{n})\mathbf{G}\,ds 
-\frac{1}{8\pi\mu}\int^{PV}_{W}(\mathbf{S}_{w}\cdot\mathbf{n})\mathbf{G}\,ds
\end{multline}
 which can be written as 
\begin{multline}\label{vel_4}
-\frac{1}{8\pi\mu}\left[\int_{S_2}((\mathbf{S}_{e}-\mathbf{S}_{i})\cdot\mathbf{n})\mathbf{G}\,ds+\int_{S_1}((\mathbf{S}_{e}-\mathbf{S}_{i})\cdot\mathbf{n})\mathbf{G}\,ds\right]
\\=\frac{1}{8\pi\mu}\int^{PV}_{W}\big(\mathbf{S}_{w}\cdot\mathbf{n}\big)\mathbf{G} \,ds
\end{multline}The computed surface temperature $T_{s1}$ and $T_{s2}$ calculates the surface tension and stress variation across the interface from \Cref{nablaS,fluid_bd_5}.The terms $(\mathbf{S}_{e}-\mathbf{S}_{i})\cdot\mathbf{n}$ and $(\mathbf{S}_{e}-\mathbf{S}_{i})\cdot\mathbf{n}$ are given in \Cref{fluid_bd_4}. From equation \Cref{vel_4} the unknown traction term $(\mathbf{S}_{w}\cdot\mathbf{n})$ can be calculated. 
Then using \Cref{vel_1,vel_2} the velocity at the interface of the droplet ($\mathbf{u}_{s1}$) and ($\mathbf{u}_{s2}$) can be calculated after taking into account the stress terms ($(\mathbf{S}_{e}-\mathbf{S}_{i})\cdot\mathbf{n}$) using \Cref{fluid_bd_4}. The velocity of a point which is not on any surface and is located within the flow field ($\mathbf{x}_0\in \mathcal{E}\cup\mathcal{I}$) is given as 
 \begin{multline}\label{vel_field}
\mathbf{u}_{if}(\mathbf{x}_0) = \mathbf{u}_{\infty}(\mathbf{x}_0)-\frac{1}{8\pi\mu}\int_{S_1}((\mathbf{S}_{e}-\mathbf{S}_{i})\cdot\mathbf{n})\mathbf{G}\,ds\\-\frac{1}{8\pi\mu}\int_{S_2}((\mathbf{S}_{e}-\mathbf{S}_{i})\cdot\mathbf{n})\mathbf{G}\,ds 
-\frac{1}{8\pi\mu}\int_{W}(\mathbf{S}_{w}\cdot\mathbf{n})\mathbf{G}\,ds
\end{multline}
The velocities at points within the flow field can be calculated for specific time steps. Thus, the calculation process requires much fewer computational field simulations using the BEM procedure. 
\par The overall thermocapillary migration is governed by hydrodynamic forces resulting from the interactions of the hydrodynamic and thermal fields of the droplets with one another and with the bounding walls, along with the surface tension lift forces that are created via the surface tension stresses on the droplet surface. The imposed flow term is just the background flow that acts relative to the droplet velocity and is a constant with time given by \Cref{vel_inf}. Understanding the effects of these interactions is of key importance in predicting the migration trajectories of the droplets. These effects can be observed from the terms in \Cref{vel_1} and \Cref{vel_2}. The second term indicates how the Marangoni stresses on one droplet affect the motion of the other. The third term describes how the wall traction modifies the trajectory of the droplet. The first term of \Cref{vel_1} is given as 
\begin{equation}\label{T_1}
\mathcal{T}_1 = \frac{1}{8\pi\mu}\int^{PV}_{S_1}((\mathbf{S}_{e}-\mathbf{S}_{i}).\mathbf{n})\mathbf{G}\,ds 
\end{equation}
 with the term $(\mathbf{S}_e-\mathbf{S}_{i}).\mathbf{n}$ being defined as
\begin{equation}\label{S_1}
(\mathbf{S}_{e}-\mathbf{S}_{i}).\mathbf{n}= Ma \nabla_S T_{s1}+\left(\frac{1}{Ca}-Ma T_{s1}\right)\mathbf{n}(\nabla.\mathbf{n})
\end{equation}
As inferred from the above equation, the part $Ma \nabla_S T_{s1}$ indicates the tangential Marangoni stress due to surface tension gradients. In contrast, the second part of this term suggests that the surface tension stress acts usually at the droplet interface. The former depends on the temperature variation on the droplet surface, with higher surface variation indicating more tangential Marangoni force. It is also dependent on the Marangoni number ($Ma$). The latter portion of this term depends on the droplet surface temperature distribution, the Capillary ($Ca$) and Marangoni numbers ($Ma$), and the droplet shape. This portion estimates the surface tension-forces developed on the droplet surface while $Ma\nabla_ST_{s1}$ denotes the Marangoni forces due to the surface temperature distribution on the first droplet. 
The second term of \Cref{vel_1} is given as 
\begin{equation}\label{T_2}
\mathcal{T}_{2} = \frac{1}{8\pi\mu}\int_{S_2}\big((\mathbf{S}_{e}-\mathbf{S}_{i})\cdot\mathbf{n}\big)\mathbf{G}\,ds 
\end{equation}
 with the term $(\mathbf{S}_{e}-\mathbf{S}_{i}).\mathbf{n}$ being defined as
\begin{equation}\label{S_2}
(\mathbf{S}_e-\mathbf{S}_{i})\cdot\mathbf{n}= Ma \nabla_S T_{s2}+\left(\frac{1}{Ca}-Ma T_{s2}\right)\mathbf{n}(\nabla\cdot\mathbf{n})
\end{equation}
Here the part $Ma \nabla_S T_{s2}$ indicates the tangential Marangoni stress acting on the first droplet due to temperature variations on the other droplet. The second part of this term suggests the stress induced by normal surface tension due to the surface temperature distribution and the shape of the second droplet, the capillary, and the Marangoni numbers. Thus, this term links how the deformation and temperature field on the second droplet's surface affect the first droplet's motion. This term indicates the droplet interaction forces. 
The third term of \Cref{vel_1} can be computed from the \Cref{vel_4}
\begin{equation}\label{T_3}
\mathcal{T}_{3} = \frac{1}{8\pi\mu}\int^{PV}_{W}(\mathbf{S}_w\cdot\mathbf{n})\mathbf{G}\,ds= -\frac{1}{8\pi\mu}\int_{S_2}\big((\mathbf{S}_e-\mathbf{S}_{i})\cdot\mathbf{n}\big)\mathbf{G}\,ds\\-\frac{1}{8\pi\mu}\int_{S_1}((S_e-S_{i})\cdot\mathbf{n})\mathbf{G}\,ds  
\end{equation}
 with the term $(\mathbf{S}_e-\mathbf{S}_{i}).\mathbf{n}$ in the first and second terms being given as  
\begin{align}\label{S_3}
(\mathbf{S}_e-\mathbf{S}_{i}).\mathbf{n}= Ma \nabla_S T_{s1}+\left(\frac{1}{Ca}-Ma T_{s1}\right)\mathbf{n}(\nabla\cdot\mathbf{n}) \\
(\mathbf{S}_e-\mathbf{S}_{i}).\mathbf{n}= Ma \nabla_S T_{s2}+\left(\frac{1}{Ca}-Ma T_{s2}\right)\mathbf{n}(\nabla \cdot \mathbf{n})
\end{align}
The above equation shows that the wall traction depends on the deformation and surface temperature distributions of the two droplets, along with the Capillary and Marangoni numbers. The contribution of this term becomes significant at higher values of confinement ratio, $Cr = 2\overline{a}/\overline{H}$. The wall traction term depends on the temperature distribution of each droplet and its deformation. The wall shear stresses compress downward force that drives the droplets toward the centreline \citep{das2018}. Thus, droplet deformation plays a crucial role in modifying the lift forces that play a part in modulating the trajectories of the droplets. 
\subsection{Parameters}

The overall dynamics of the thermocapillary flow are usually dependent on the Capillary number of the flow ($Ca$) and also on the initial drop separation ($\mathbf{q}_0$) along with the Marangoni number ($Ma$) and the confinement ratio ($Cr$). Droplets close to each other and placed in narrowly confined channels undergo a higher degree of deformation than those placed far apart and in wider channels. The Marangoni number ($Ma$) indicates the strength of the thermal Marangoni forces that act tangentially to the droplet surfaces, and higher Marangoni numbers can also lead to higher droplet deformations. The confinement ratio ($Cr$) and initial droplet separation($\mathbf{q}_{0}$) determine the relative importance of the latter two terms in \Cref{T_2} and \Cref{T_3}, respectively. Experiments have shown that thermal Marangoni numbers ($Ma$) can be varied from $1-5$ in a microfluidic system as in \citet{Santra2023}. The capillary numbers are kept small to avoid the effects of large droplet deformation. Except for the last section, the drops have been placed at equal distances from the wall ($e_l=0.1,p_l=0$).
\subsection{Discretisation of the space and time}
\subsubsection{Mesh generation}
The boundary integral \label{temp_1,temp_2,temp_3} for the interfacial temperature and the velocity \label{vel_1,vel_2,vel_4} can be solved on a dynamically evolving unstructured grid. The grid generation procedure is outlined in \citep{yon1998,pozrikidis2002practical} and has been adapted from the open-source package BEMLIB. Curved 6-node triangular elements are used with three vertex nodes and three midpoint nodes. In our study, we used 2562 collocation nodes ($\mathbf{x}_{0}$) on each droplet ($S_{l}$) and 2704 nodes on the wall surface (W).
\subsubsection{Evaluation of singular and non-singular integrals}
The curved 6-node triangular elements are mapped onto a 2D parametric space via a second-order iso-parametric mapping. This form of mapping uses the same interpolation function for the geometric and the field variables of interest to project the curved element onto the 2D parametric space, taking the shape of a unit right angle. When the collocation point lies on the element, it is called a singular element. For any singular element, the entire element is divided into 4 parts, and a polar integration rule outlined in  \citep{yon1998,pozrikidis2002} is used to compute the single layer integral in each part. The double-layer integrals are obtained by removing the singular part from the main integral and then evaluating the integral. The integral over the singular portion is then computed analytically and added.. The single and double-layer integrals are calculated using a 7-point Gauss-Legendre quadrature for non-singular elements. For a detailed discussion on the calculation of the integrals, we refer the reader to   \citet{pozrikidis2002practical}.  
\subsubsection{Computation of the stress tensor discontinuity}
The stress tensor discontinuities \Cref{S_3} are computed using the surface tension values, its gradient and the unit normals and curvatures at the points. The technique of \cite{zinchenko1997} is used to calculate the unit normals and curvatures. The interfacial tension is calculated using \Cref{fluid_bd_4}. The surface tension gradient is computed using the surface tension in the Cartesian grid and is then mapped to the local parametric space of the particular element \citep{yon1998}. 
\subsubsection{Interface Tracking}
The interface is tracked using a second-order Runge-Kutta scheme given as 
\begin{equation}
    \mathbf{x}_{n+1} = \mathbf{x}_n+\frac{\delta t}{2}(\mathbf{u}_n+\mathbf{u}_{n+\frac{1}{2}})
\end{equation}
Here the terms $\mathbf{u}_n$ and $\mathbf{u}_{n+\frac{1}{2}}$ represent the velocities of the marker points at the $n^{th}$ time step and the intermediate time step computed from the values of the position and velocity of the interface at the $n^{th}$ step while $\delta t$ is the time step value. The velocity $\mathbf{u}_n$ is the superposition of the normal velocity at the interface marker points and the relaxation velocity of the mesh. This Mesh relaxation velocity is added to ensure the stability of the mesh at high levels of deformation and does not interfere with the underlying dynamics of the flow. The mesh relaxation velocity is obtained from minimising the mesh distortion energy and has been adapted from \cite{zinchenko1997}.
\subsubsection{Post processing}
The droplet centre velocity ($\mathbf{u}_c$) is computed from the area average of the velocities over the droplet surface ($S_{l}$), and this corresponds to the velocity of the droplet centre. The droplet centre is computed by taking the area average of all the coordinates of points on the surface of the droplet surface ($S_{l}$). The droplet centre positions $\mathbf{x}_c$ are plotted to give the migration trajectories of the droplets. The velocity of points in the domain ($\mathbf{u}_{if}$) are calculated using \Cref{vel_field}. The field velocity data is visualized using commercial software, Paraview. 
\section{Results and discussion}
We have performed the three-dimensional boundary element computations of the thermocapillary migration of two droplets surrounded by non-slip walls. However, before discussing the results, we present the results for the droplets in a confined isothermal flow. After that, we imposed a thermal gradient within the confined flow domain to observe the thermocapillary effects on the droplet migration pattern. 
Without the temperature gradient, we observe a transverse droplet migration towards the wall, similar to the observation made by \cite{Mortaza2002} for a single droplet.
However, the presence of another droplet significantly alters the hydrodynamic and thermal fields. This alters the droplets' temperature and deformation, causing variations in the thermal Marangoni and hydrodynamically-induced lift forces. The droplet deformation and migration change the fluid hydrodynamics. In return, the fluid hydrodynamics also alters the droplet behaviour and migration patterns via droplet-droplet interactions. This can lead to significant deviations in the migration patterns of multiple confined droplets from the case of a single droplet in an unbounded domain. We seek to understand how the migrations of these droplets are affected by the domain confinement ($Cr$) and the initial $x$-separation of the two droplets ($d_{0}$). More importantly, it is necessary to understand the regions where the droplet migration is solely influenced by its induced thermal field rather than interactions with other droplets in the vicinity. Since the interactions with the other droplets happen via the continuous medium, this analysis will help us understand when the droplet migration can be more actively controlled. We can also understand the zones within which the surface tension-driven motion happens as opposed to the motion brought about by the droplet interaction-induced and wall-induced lift forces. 
\subsection{Flow in an isothermal domain}
\begin{figure}
    \centering
    \includegraphics[width=0.7\textwidth]{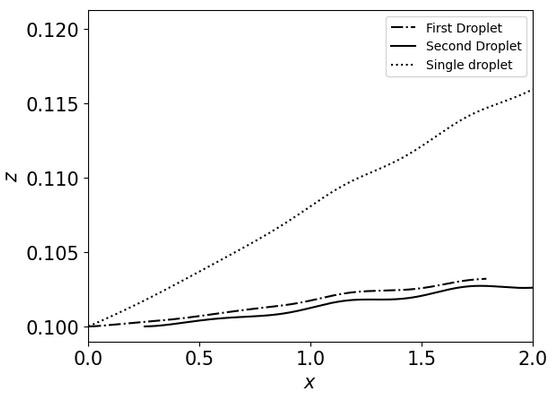}
    \vspace{-0.1in}
    \centering\caption{Centre trajectories of two drops in contact in the absence of temperature gradient ($Cr=0.25$, $d_{0}=0.25$), compared to the migration of a single droplet \citep{Mortaza2002}.}
    \label{zvx_iso_drops}
\end{figure}

\begin{figure}
\centering
\minipage{0.48\textwidth}
\includegraphics[clip,width =1.0\textwidth]{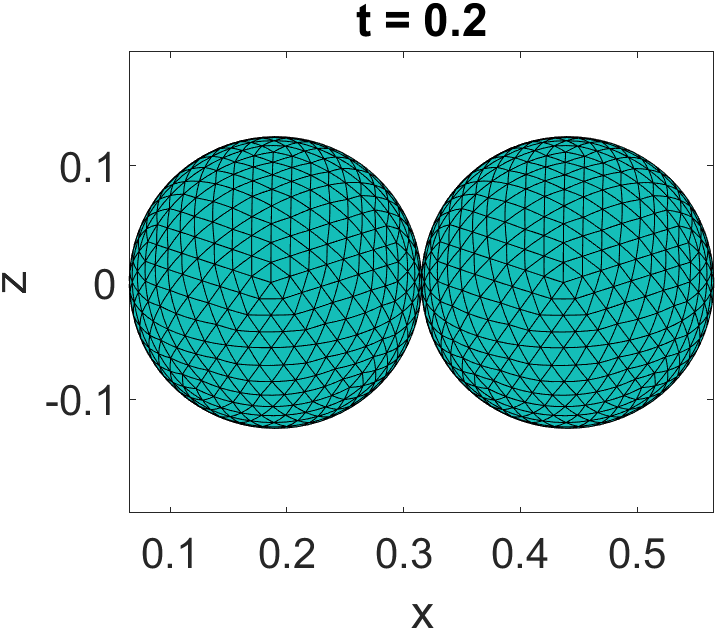}
\endminipage
\minipage{0.48\textwidth}
\includegraphics[clip,width =1.0\textwidth]{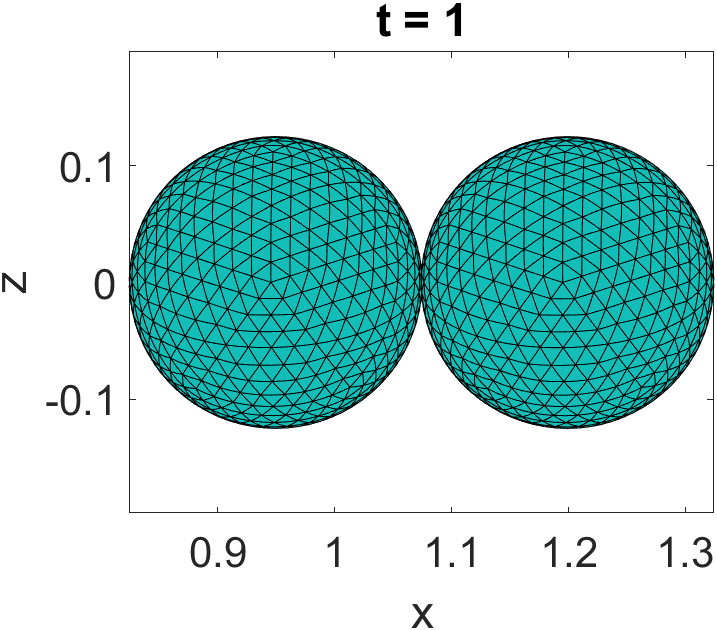}
\endminipage
\caption{Translation of migrating droplets, initially touching, in an isothermal environment.}
\label{notemp_t}
\begin{picture}(1,1)
\setlength{\unitlength}{1cm}
\end{picture}
\end{figure}

Two droplets of equal radius and viscosity of the constituent fluid are positioned to touch each other at an offset distance ($e_{1}=e_{2}=0.1$, $p_{1} = p_{2} = 0,$ \Cref{setup}). Their migration characteristics are studied, assuming isothermal conditions. Presently, two droplets are considered, wherein the trailing droplet is placed at the channel inlet. The droplet separation vector is given as $\mathbf{q}_{0} = \{0.25,0,0\}$, and the confinement ratio of $Cr = 0.25$. The capillary number ($Ca$) is taken to be 0.25. Without the temperature gradient, the forces acting on the droplet are due to the walls, the droplet interactions, and surface tension forces. The effect of the wall is very low for this case due to the small confinement, $Cr=0.25$ \citep{keh2002}. The migration trajectories of droplets in the absence of a temperature gradient are shown in \Cref{zvx_iso_drops}. 
 The upward transverse migration towards the wall observed in \Cref{zvx_iso_drops} can be explained by the normal forces manifesting as pressure gradients induced by the surface tension forces acting on the droplet surface. This is what we see for a single droplet as well \citep{Mortaza2002}; however, the migration for two droplets is less pronounced due to the droplet interaction forces. As seen in \Cref{zvx_iso_drops}, the single droplet rises faster towards the wall than the multiple droplets. This shows that the droplet deformation due to droplet interactions pushes the droplets downward, thus causing a slower upward motion than in the single droplet case. In the quasi-steady limit of low Reynolds and Peclet numbers, droplets of equal size flow together as one unit as depicted in \Cref{notemp_t}. 

\subsection{Flow in a non-isothermal domain}
 To observe the thermocapillary behaviour of the droplets, an axial temperature gradient along the longitudinal direction is imposed as in \cref{heat_bd_2}. The side walls are considered to be insulated (\cref{heat_bd_3}). Without any convection due to the low Peclet numbers, a linearly decreasing axial temperature gradient is present in the imposed flow that interacts with the droplets. The droplet separation is $\mathbf{q}_{0} = \{0.25,0,0\}$, and a confinement ratio of $Cr = 0.25$ and a conductivity ratio of $\delta = 0.1$ is assumed. The temperature gradient induces variations of surface tension gradients on the droplet surface, thereby causing surface tension-induced lift forces that change the migration dynamics of the droplets. The interaction forces between the droplets are also altered due to the presence of the thermal field.  
   \begin{figure}
    \centering
    \includegraphics[width=0.7\textwidth]{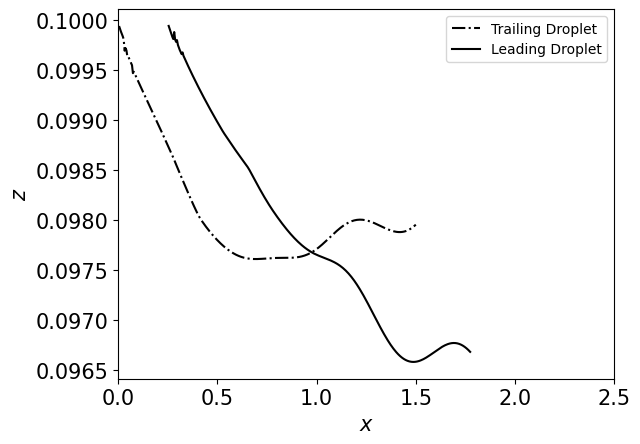}
    \vspace{-0.1in}
    \centering\caption{Droplet center trajectories of two drops in contact in the presence of temperature gradient ($Cr=0.25$, $d_{0}=0.25$)}
    \label{zvx_non_iso_drops}
\end{figure} 
With the temperature gradient, we observe a transverse droplet migration away from the wall and towards the channel centreline (\Cref{zvx_non_iso_drops}) as opposed to the transverse migration towards the wall in the case of isothermal flow seen in \Cref{zvx_iso_drops}. The downward transverse migration away from the wall can be explained by the altered lift forces due to changes in the surface tension stresses due to the Marangoni effect. The droplet interaction forces between the droplets are also changed since the surface tension gradients are altered for both the droplets. With the addition of a temperature gradient within the flow, we obtain a gradual separation of the droplets depicted in \Cref{temp_t}. It can be stated that the droplets separate due to varying thermal gradients between themselves since one droplet is exposed to different temperatures, leading to different Marangoni forces. 
\begin{figure}
\centering
\minipage{0.48\textwidth}
\includegraphics[clip,width =1.0\textwidth]{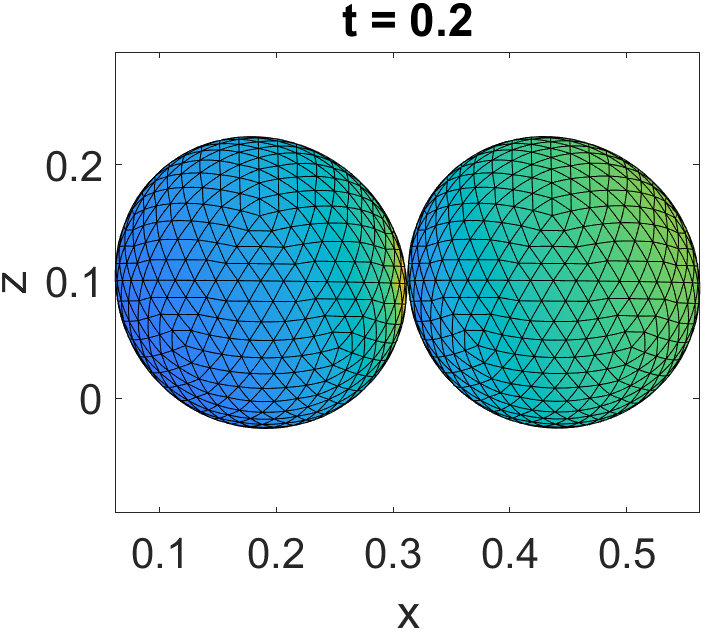}
\endminipage
\minipage{0.48\textwidth}
\includegraphics[clip,width =1.0\textwidth]{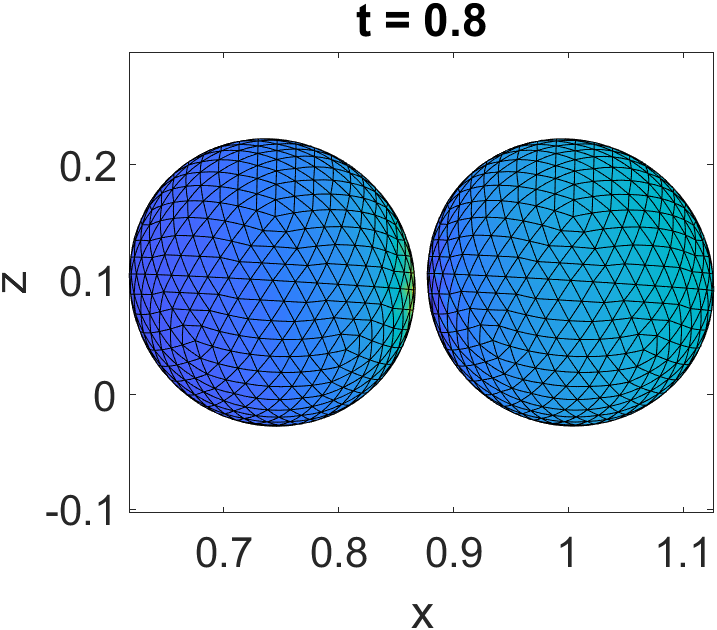}
\endminipage
\caption{Translation of migrating droplets, initially touching, in a non-isothermal environment.}
\label{temp_t}
\begin{picture}(1,1)
\setlength{\unitlength}{1cm}
\end{picture}
\end{figure}

\subsection{Effect of confinement ratio on the droplet migration trajectory}
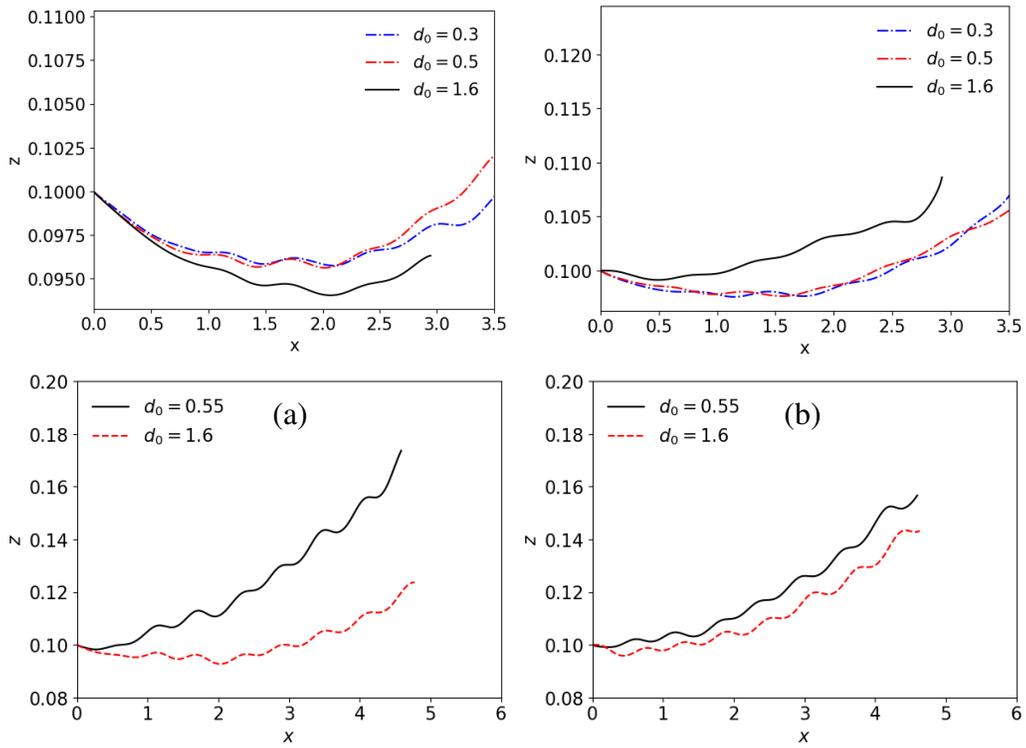
\begin{figure}
    \centering
    \minipage{0.5\textwidth}
    \includegraphics[clip,trim = 0pt 0pt 0pt 0pt,width=\linewidth]{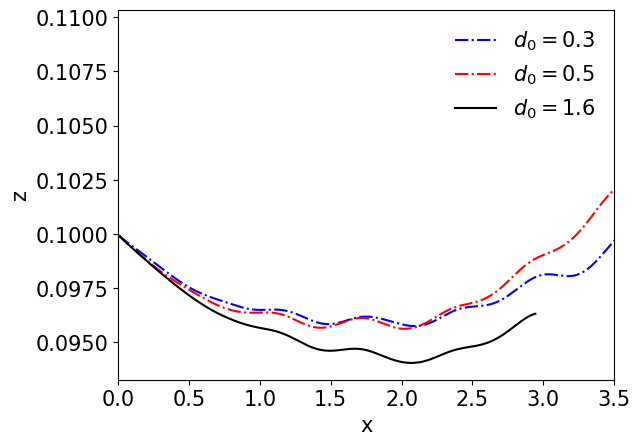}
    \endminipage\hfill
    \minipage{0.5\textwidth}
    \includegraphics[clip,trim = 0pt 0pt 0pt 0pt,width=\linewidth]{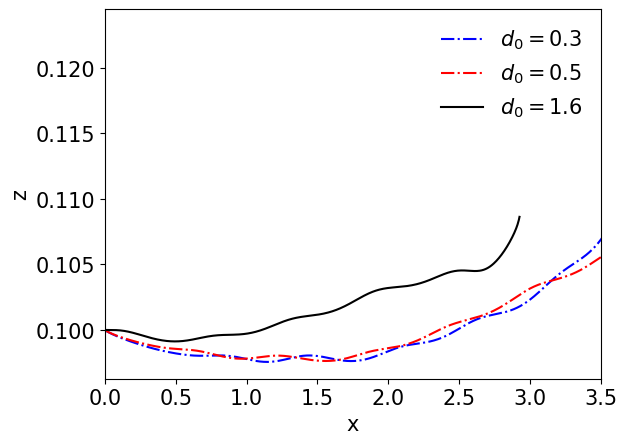}
    \endminipage

    \minipage{0.5\textwidth}
    \includegraphics[clip,trim = 0pt 0pt 0pt 0pt,width=\linewidth]{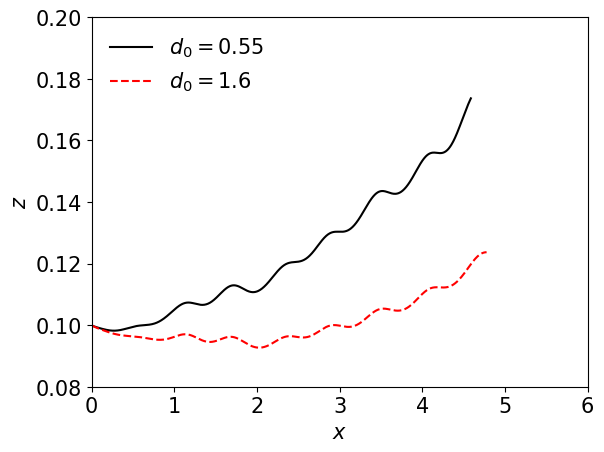}
    \endminipage\hfill
    \minipage{0.5\textwidth}
    \includegraphics[clip,trim = 0pt 0pt 0pt 0pt,width=\linewidth]{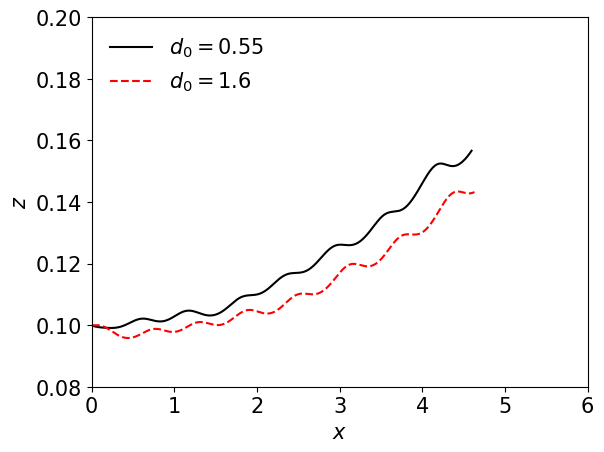}
    \endminipage

    \caption{Variation of droplet centre trajectories with initial droplet separation ($d_0$) for $Cr=0.25$ (top), $Cr=0.5$ (bottom). Left column: trailing droplet, right column: leading droplet.}
    \label{z_v_x_d0}
\begin{picture}(1,1)
\setlength{\unitlength}{1cm}
\put(-3.2,6.6){(a)}
\put(3.6,6.6){(b)}
\put(-3.2,1.5){(c)}
\put(3.6,1.5){(d)}
\end{picture}
\end{figure}

We consider the effect of the confinement of the domain ($Cr$) on the migration of the droplets for varying initial $x$-separation ($d_{0}$) between them. Other parameters ($e_{1}=e_{2}=0.1$,$p_{1}=p_{2}=0$), ($Ca=0.25$, $Ma=0.5$) and ($\delta = 0.1$) are kept unchanged. The wall-induced forces are negligible for a smaller confinement ratio; therefore, thermal interactions only induce droplet deformation. Computations are performed for a confinement ratio $Cr=0.25$, and for $d_{0}$ of $0.3$, $0.35$, and $1.6$.
For $Cr = 0.2$ or lower, the droplet behaviour is similar to that in an unbounded channel \citep{keh2002}. Thus, with $Cr = 0.25$, the droplet is in an almost unbounded medium. The droplet migration trajectories are observed and plotted in \Cref{z_v_x_d0}(a),(b) for the two droplets in two columns. The droplets are shown to move towards the drop centreline and then change direction towards the channel wall. This trajectory moves downwards for the trailing droplet (\Cref{z_v_x_d0}(a)) as the droplet-separation distance ($d_{0}$) increases. In the case of the leading droplet (\Cref{z_v_x_d0}(b)), this trajectory transitions to a monotonically increasing curve at higher separation distances. It is observed that the trailing droplet tends to push more downward, whereas the leading droplet tends to go upward more rapidly toward the wall when they are placed close to each other. 
\par Subsequently, the confinement is increased to $Cr = 0.5$. For this degree of confinement in the domain, some alterations in the thermal and hydrodynamic fields from an almost unbounded case ($Cr=0.25$) are expected. The initial separation distance($d_{0}$) between the droplets is varied to assess the interactions between their thermal and hydrodynamic fields while keeping all geometric ($e_{l},p_{l}$) and physical ($Ca, Ma$) parameters the same as in the previous case. 
The overall migration trajectory has changed from a parabolic nature to a monotonically increasing curve as is depicted in \Cref{z_v_x_d0}(c),(d). The monotonically increasing upward migration of the droplets reduces with separation distance ($d_{0}$) for both the trailing and the leading droplet. For large distances, it is reduced to the extent to which the monotonically increasing curve flattens to a general parabolic nature. Droplet migration is less affected by initial separation ($d_{0}$) for the leading droplet than the trailing drop. 
\subsection{Understanding the physics of droplet migration}
\par As the droplets migrate along the channel through the decreasing axial temperature field, the leading side of the faces is exposed to a lower temperature, while the trailing side is at a higher temperature. As a result, we observe a variation in the surface tension due to \cref{fluid_bd_4} across the surface of both droplets, leading to the development of surface tension forces acting tangentially to the droplet interface. These surface tension forces are one of the dictating factors in governing droplet trajectories, along with wall-induced lift forces and the interaction forces between the droplets. The velocity of the imposed flow ($\mathbf{u}_{\infty}$) only causes the axial migration of the droplets. In contrast, the transverse migration of the droplets depends on the Marangoni stresses due to the surface tension gradients, the interaction of hydrodynamic and thermal fields between the two droplets, and the presence of bounding walls. To understand the controllability of the motion of each droplet, we focus our attention on the effect of the presence of the other droplet and the bounding walls.
If we look at \cref{vel_1} and \cref{vel_2}, the velocities of the two droplets are determined by the summation of three separate vectors. Out of the three, the term given in \cref{T_1} corresponds to the droplet surface tension forces and its influence on the droplet motion trajectory, while the other two terms given by \cref{T_2} and \cref{T_3} refer to the effect of the droplet interaction forces and the wall-induced forces, respectively. To define their relative influence on the droplet, we define the influence parameter $S$ as 
\begin{equation}\label{influence}
    S = \left|\frac{\mathbf{T}^{avg}_1\cdot(\mathbf{U}_c-\mathbf{U}_{\infty,c})}{|\mathbf{U}_c-\mathbf{U}_{\infty,c}|^2} -1\right|
\end{equation}
where
\begin{equation}
    \mathbf{T}^{avg}_1 = \frac{\int_{s_{l}}\mathcal{T}_1\,ds}{A_{s_{l}}}
\end{equation}
is the area average of the term in \cref{T_1} over the droplet surface ($S_{l}$). The imposed flow velocity($u_{\infty,c}$) must be subtracted to capture the physics of the droplet relative to the background flow; here, $u_{\infty,c}$ is the imposed flow velocity of the droplet centre. The influence parameter $S$ refers to surface tension forces and their overall contribution to the velocity of the droplet. The parameter $S\sim 0$ if the Marangoni forces are more important than the interaction and wall-induced forces. On the other hand, if $S>>0$, the surface temperature distribution on the other droplet and wall-induced lift forces play a more critical role in determining the droplet motion.  In the former case, the droplet motion is driven mainly by the thermal field that develops on its surface due to conduction from the axially imposed temperature field via the induced surface tension forces. The latter indicates that the combined role of the droplet interaction and wall-induced forces is high. The variation in $S$ with time is depicted in \Cref{stgrad_v_x}. The wall-induced lift terms are almost negligible for droplets with lower confinement ($Cr = 0.25$). Since the influence parameters are mostly positive and range from $1$ to $1.5$, it can be stated that both surface tension and droplet-interaction forces drive the droplet motion. There are certain instances when the droplet surface tension forces have a considerably higher influence, shown by the minima of the curve (\Cref{stgrad_v_x}(a),(b)) for both the leading and trailing droplets, especially for $d_{0}=0.3,0.5$, immediately followed by high values of the influence coefficient. Thus, the dominance of the thermocapillary forces at one instance is followed by the dominance of the droplet interaction forces in the next. This leads to the oscillatory nature in the variations of $S$ with $t$. It is also seen from \Cref{stgrad_v_x}(b) that, overall, the influence parameter ($S$) fluctuates more with time ($t$) for the leading droplet. The mean value of the leading droplet influence parameter($S$) decreases with time for all separation distances while remaining constant for the trailing droplet, as shown in \Cref{stgrad_v_x}(a),(b). The leading droplet moves in the imposed flow, while the trailing droplet is immersed within the wake of the leading droplet. This might explain why the mean values of $S$ for the leading droplet are lower compared to the trailing droplet. The overall mean value of the parameter ($S$) is lower for $d_0=1.6$, decreasing to a steady value close to zero for the leading droplet (\Cref{stgrad_v_x}(b)). Thus, it can be stated that with time, the motion of the leading droplet for $d_0=1.6$ becomes entirely independent of the presence of the trailing droplet.   
Overall, the mean value of the droplet influence terms for $d_0=1.6$ is the lowest for the leading droplet and highest for the trailing droplet. The droplet influence terms for $d_{0} = 0.3,0.5$ are generally similar for the trailing droplet, while for the leading droplet, the $S$ term for $d_{0}=0.5$ is slightly lower compared to $d_0=0.3$. 
\par When the confinement ratio is increased, the values of the influence terms become stable for both the leading and trailing droplets and develop a constant value close to $1.0$ as shown in \Cref{stgrad_v_x}(c),(d). The gradual decrease in influence parameters with time for the leading droplet seen at lower confinements (\Cref{stgrad_v_x}(b)) is absent. Instead, the influence coefficients remain mostly steady with time, with a few oscillations present. The reduction in oscillations of $S$ with time might be due to increased wall-induced forces and higher droplet interactions. The wall effects are higher for higher confinements, leading to more wall-induced hydrodynamic forces affecting the droplet migration. From \Cref{stgrad_v_x_contour}(a), we can see that there are also increased droplet interactions in this case. Thus, unlike what we see for lower confinements where the surface tension and interaction forces dominate alternately, there are no instances when the surface tension force dominates. The equal influence of all three forces on the droplet motion gives rise to the stable value of $S$ depicted in \Cref{stgrad_v_x}(c)(d). The mean values of the influence terms for the trailing droplet and the leading droplet reduce marginally with separation distance ($d_{0} = 0.55,1.6$) due to the reduced droplet interactions. 
The surface tension and droplet-interaction forces prevail at lower confinement ratios, while the wall-induced forces affect the motion at higher confinements. However, in some regimes of the droplet migration trajectory, the droplet motion has a higher surface tension influence relative to other forces, even at higher confinements. These are minima in the $S$ vs $t$ curve for the leading droplet at $Cr = 0.5,d_{0}=1.6$ shown in \Cref{stgrad_v_x}(d). 
\begin{figure}
    \centering
    \minipage{0.24\textwidth}
    \includegraphics[clip,trim = 0pt 0pt 0pt 0pt,width=\linewidth]{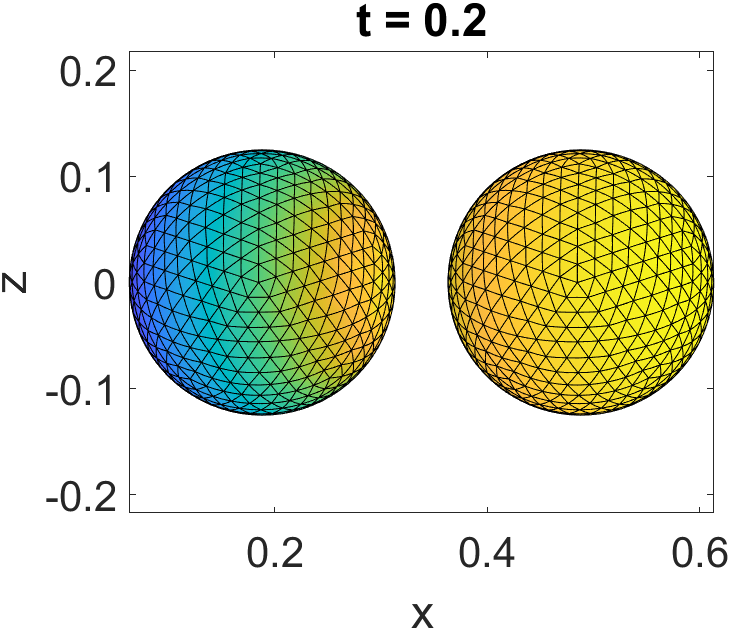}
    \endminipage
    \minipage{0.24\textwidth}
    \includegraphics[clip,trim = 12pt 0pt 0pt 0pt,width=\linewidth]{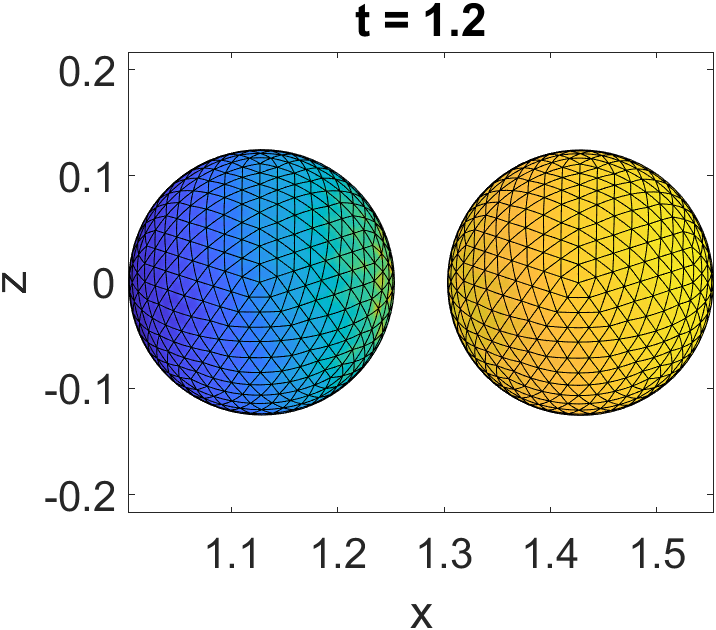}
    \endminipage
    \minipage{0.24\textwidth}
    \includegraphics[clip,trim = 12pt 0pt 0pt 0pt,width=\linewidth]{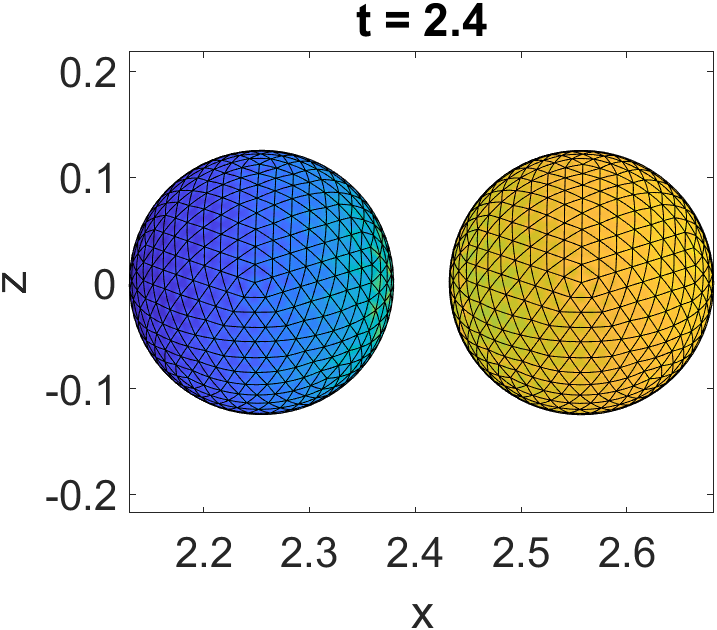}
    \endminipage
    \minipage{0.24\textwidth}
    \includegraphics[clip,trim = 12pt 0pt 0pt 0pt,width=\linewidth]{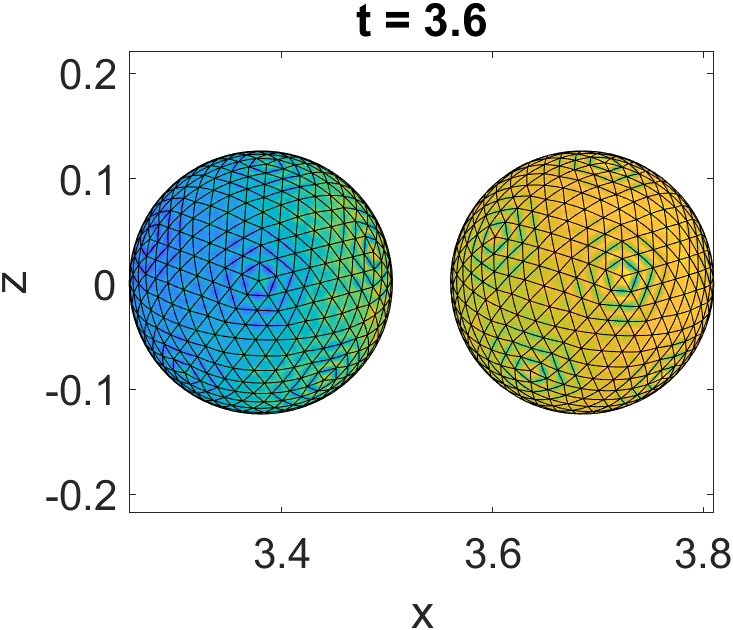}
    \endminipage
    \\
    \minipage{0.24\textwidth}
    \includegraphics[clip,trim = 0pt 0pt 0pt 18pt,width=\linewidth]{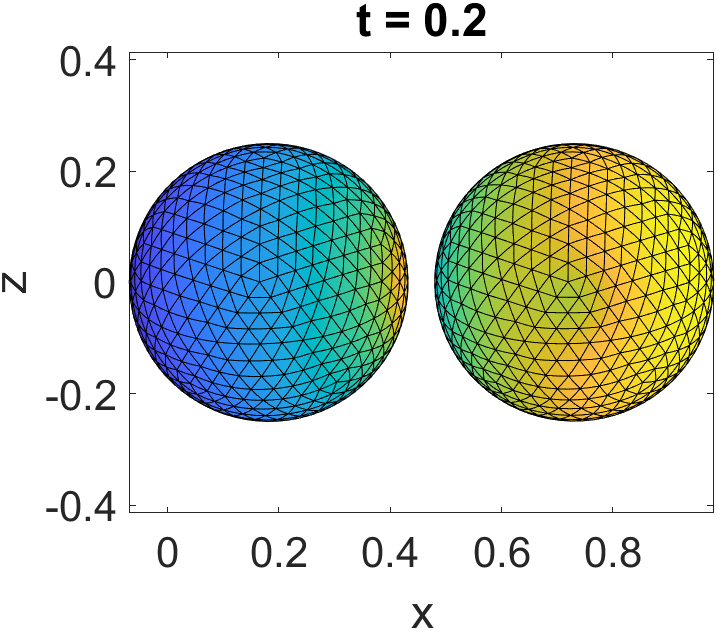}
    \endminipage
    \minipage{0.24\textwidth}
    \includegraphics[clip,trim = 12pt 0pt 0pt 18pt,width=\linewidth]{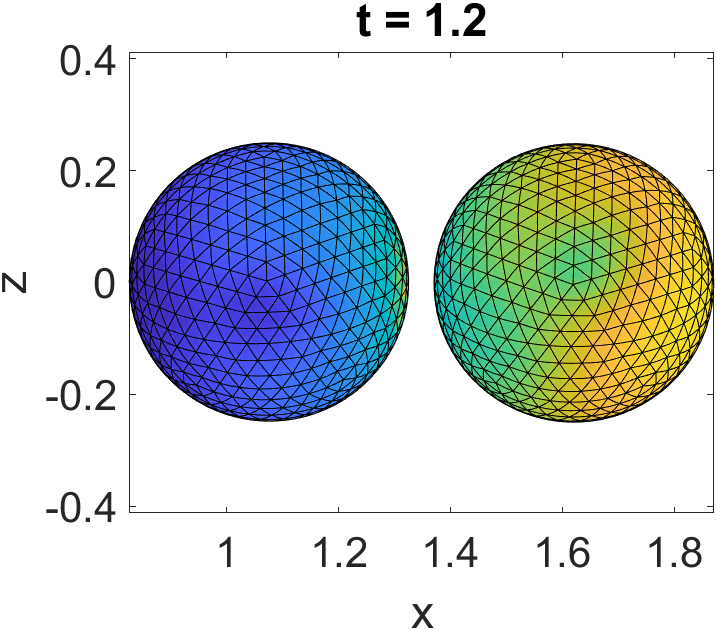}
    \endminipage
    \minipage{0.24\textwidth}
    \includegraphics[clip,trim = 12pt 0pt 0pt 18pt,width=\linewidth]{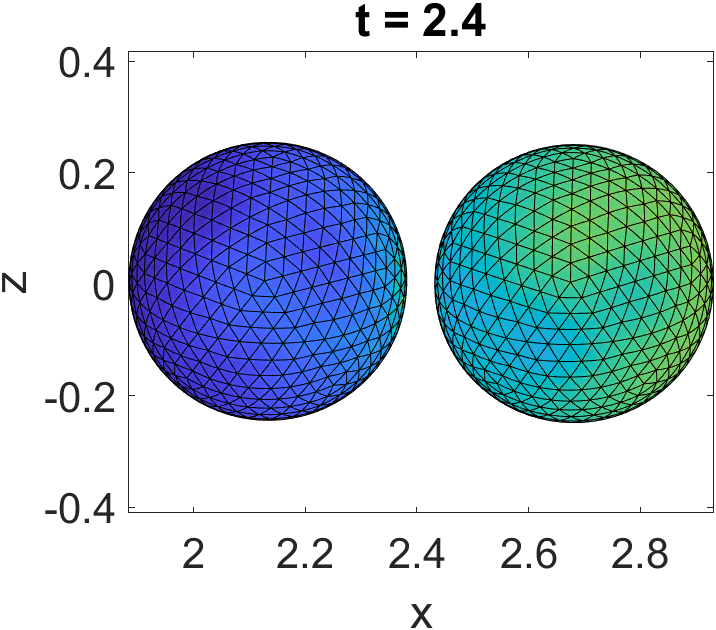}
    \endminipage
    \minipage{0.24\textwidth}
    \includegraphics[clip,trim = 12pt 0pt 0pt 18pt,width=\linewidth]{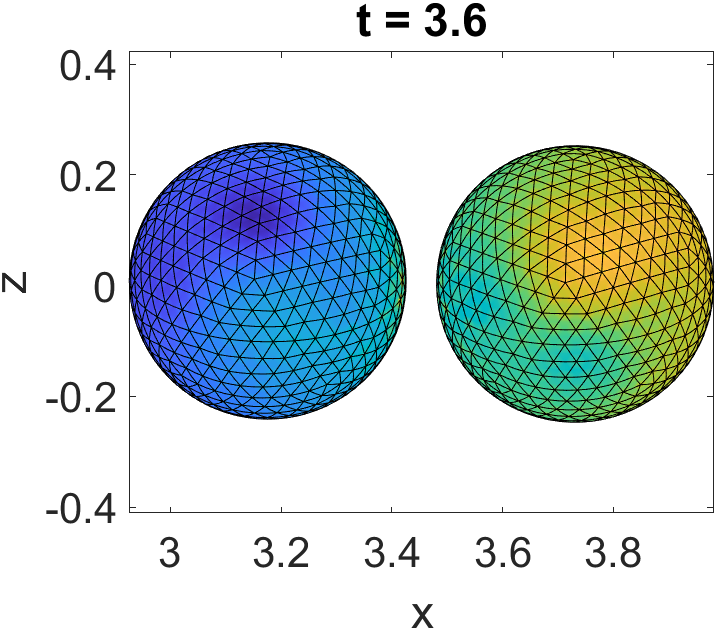}
    \endminipage
    \caption{Evolution of droplet surface tension gradients for different confinement ratios. Left: Trailing droplet right: Leading droplet Top: ($Cr=0.25$), bottom: ($Cr=0.5$)}
    \label{drop_evol_st}
\end{figure}
\begin{figure}
    \centering
    \minipage{0.5\textwidth}
    \includegraphics[clip,trim = 20pt 0pt 0pt 5pt,width=\linewidth]{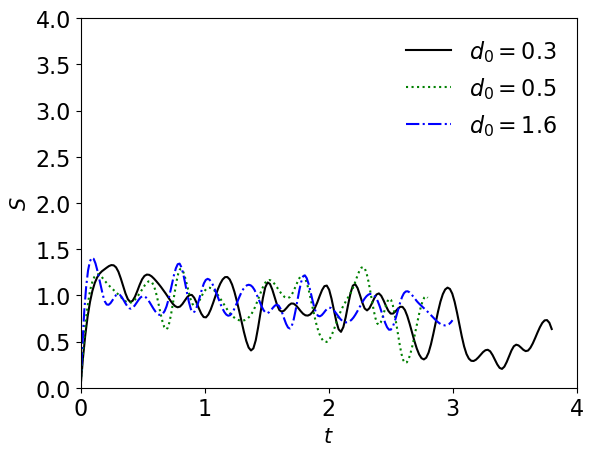}
    \endminipage\hfill
    \minipage{0.5\textwidth}
    \includegraphics[clip,trim = 25pt 0pt 0pt 0pt,width=\linewidth]{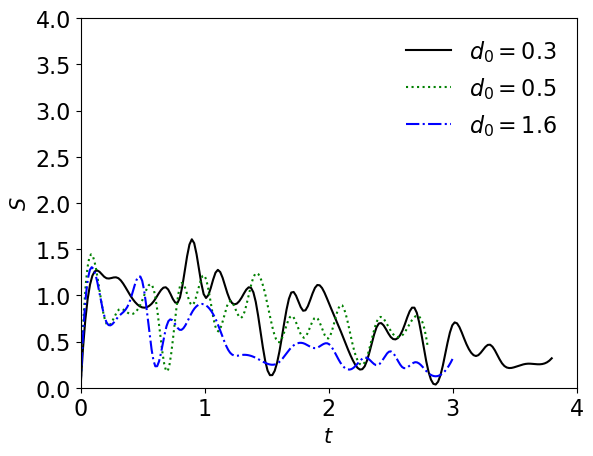}
    \endminipage
    
    \minipage{0.5\textwidth}
    \includegraphics[clip,trim = 20pt 0pt 0pt 0pt,width=\linewidth]{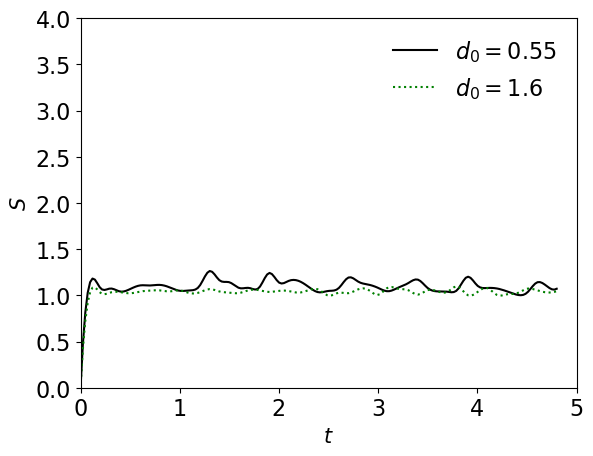}
    \endminipage\hfill
    \minipage{0.5\textwidth}
    \includegraphics[clip,trim = 25pt 0pt 0pt 0pt,width=\linewidth]{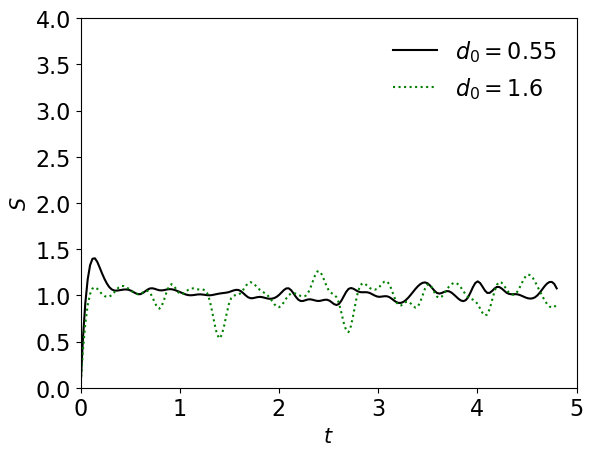}
    \endminipage
    
    \caption{Variation of droplet influence parameter (S) with initial droplet separation for $Cr=0.25$ (top), $Cr=0.5$ (bottom). Left column: trailing droplet, right column: leading droplet.}
    \label{stgrad_v_x}
\begin{picture}(1,1)
\setlength{\unitlength}{1cm}
\put(-3.2,7.1){(a)}
\put(3.6,7.1){(b)}
\put(-3.2,1.5){(c)}
\put(3.6,1.5){(d)}
\put(-7.2,5){\rotatebox{90}S}
\put(-7.2,10.5){\rotatebox{90}S}
\end{picture}    
\end{figure}
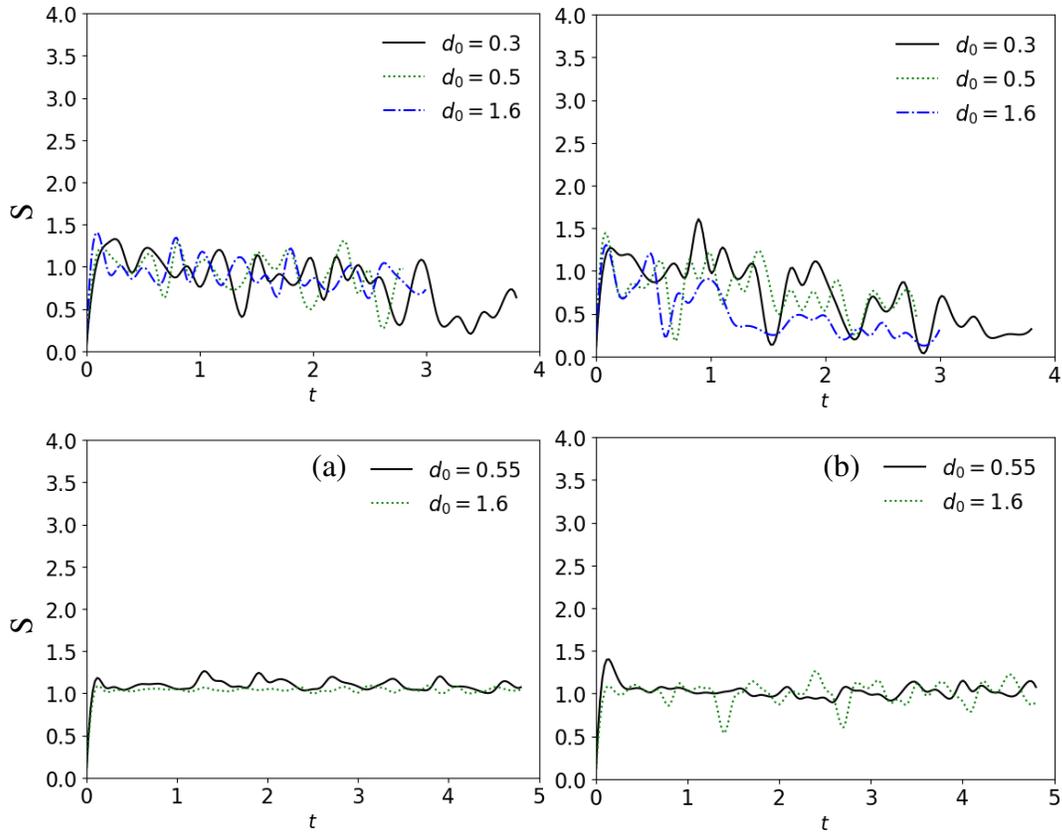
\par The velocity contours for both confinement ratios ($Cr$) and a particular value of $d_0$ are depicted in \Cref{stgrad_v_x_contour}. For both the confined cases, there is a velocity gradient between the inside and the outside of the droplet. However, in the higher confined cases \Cref{stgrad_v_x_contour}(a), the velocity contours along the lower surface of the two droplets are connected to one another via a bridge, forming a zone of interaction. Fluid from the upper and lower surfaces of the two droplets mix along this zone, leading to similar velocities within this zone. This bridge becomes thicker with time. Thus, we see strong droplet interactions in the higher confined case, which is not what we see for the lower confined case. In the lower confined case, as seen from \Cref{stgrad_v_x_contour}(b), the envelope of similar velocities covering the two droplets' upper and lower parts are absent except in the initial motion stages. The bridge-like structure is present initially for the lower confined case but gradually disappears with time. The differences between the velocities within the droplet and the outside flow are low for both confinements. However, strong droplet interactions exist between the two droplets in the higher confined case depicted in \Cref{stgrad_v_x_contour}(a). This might be due to higher confinement; the upper and lower surfaces of the droplet reach the boundary layer regions close to the wall. Due to this, the interior of the droplets, which have velocities similar to the reduced velocity regions close to the wall, can form a zone of interaction that surrounds both droplets.  
\begin{figure}
    \centering
    \minipage{0.48\textwidth}
    \includegraphics[clip,trim = 0pt 0pt 0pt 60pt,width=\linewidth]{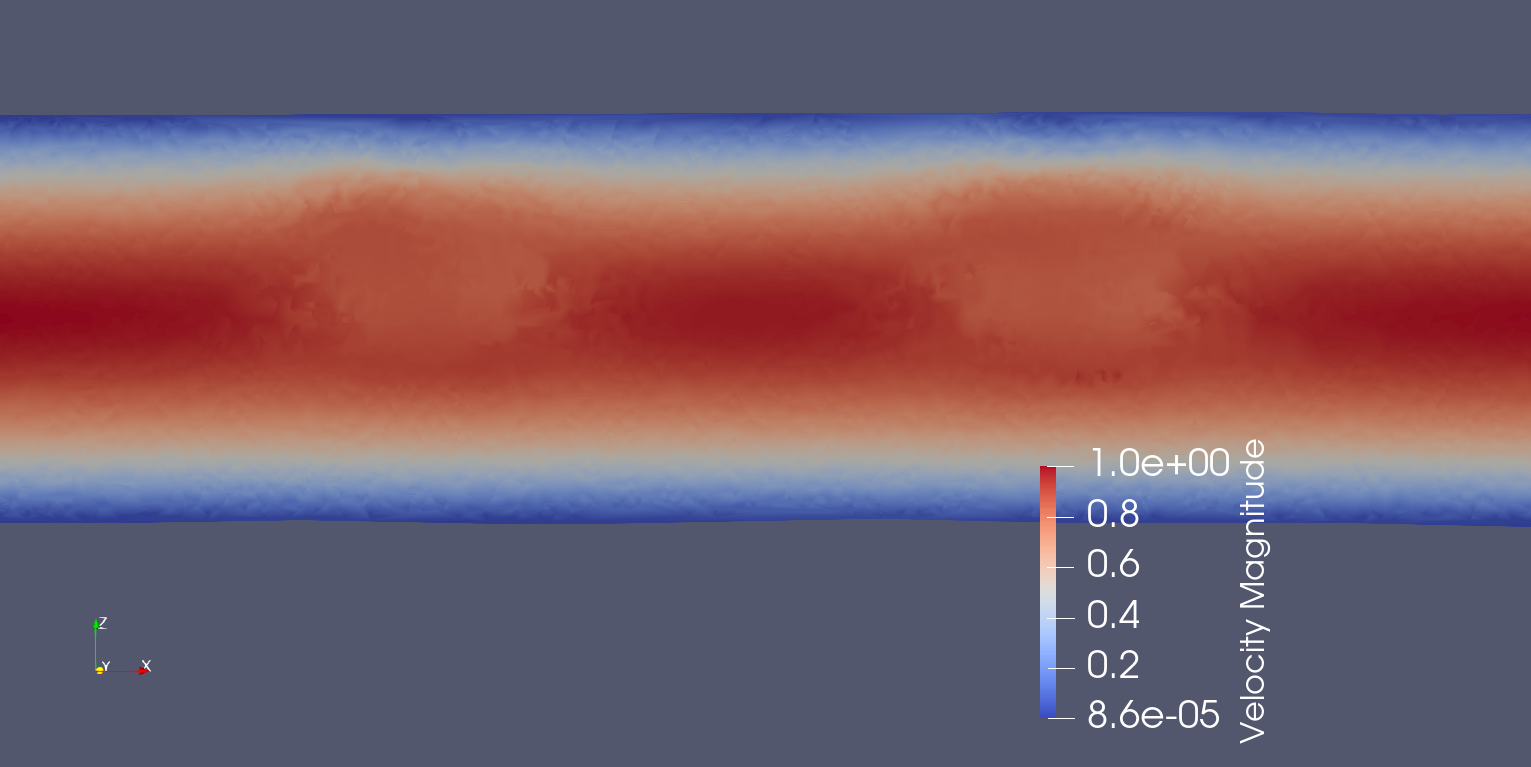}
    \endminipage
    \minipage{0.48\textwidth}
    \includegraphics[clip,trim = 0pt 0pt 0pt 60pt,width=\linewidth]{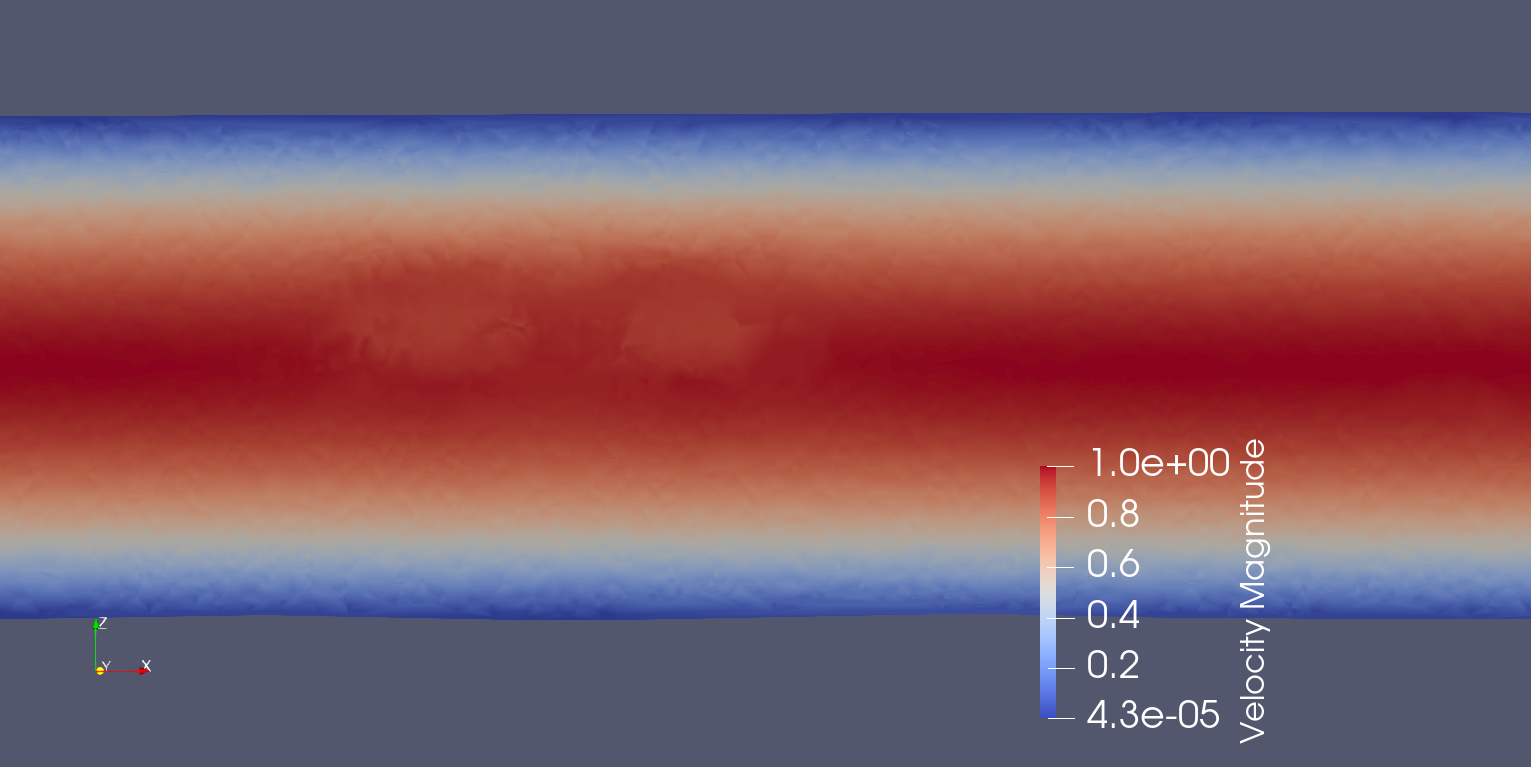}
    \endminipage
    
    \minipage{0.48\textwidth}
    \includegraphics[clip,trim = 0pt 0pt 0pt 60pt,width=\linewidth]{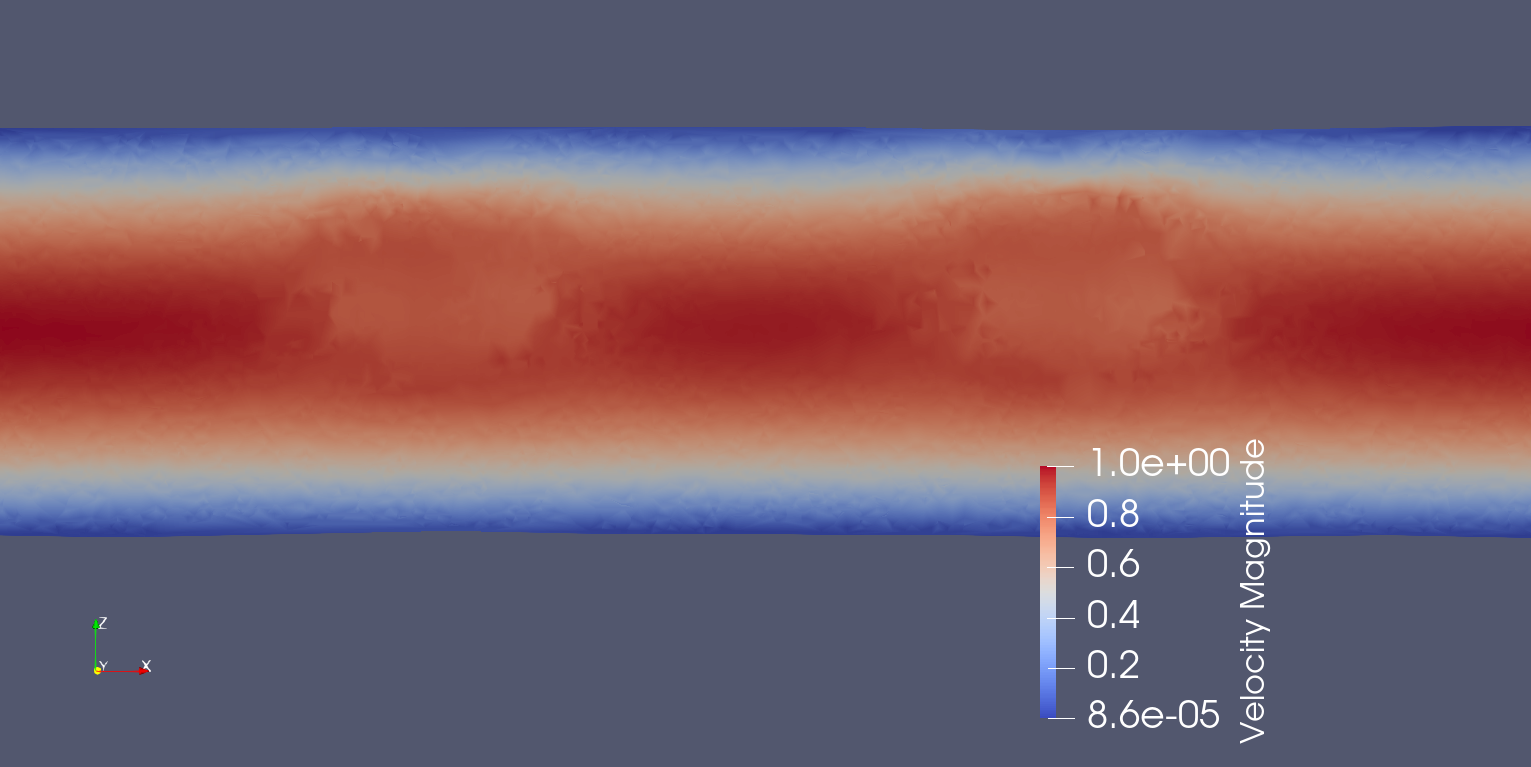}
    \endminipage
    \minipage{0.48\textwidth}
    \includegraphics[clip,trim = 0pt 0pt 0pt 60pt,width=\linewidth]{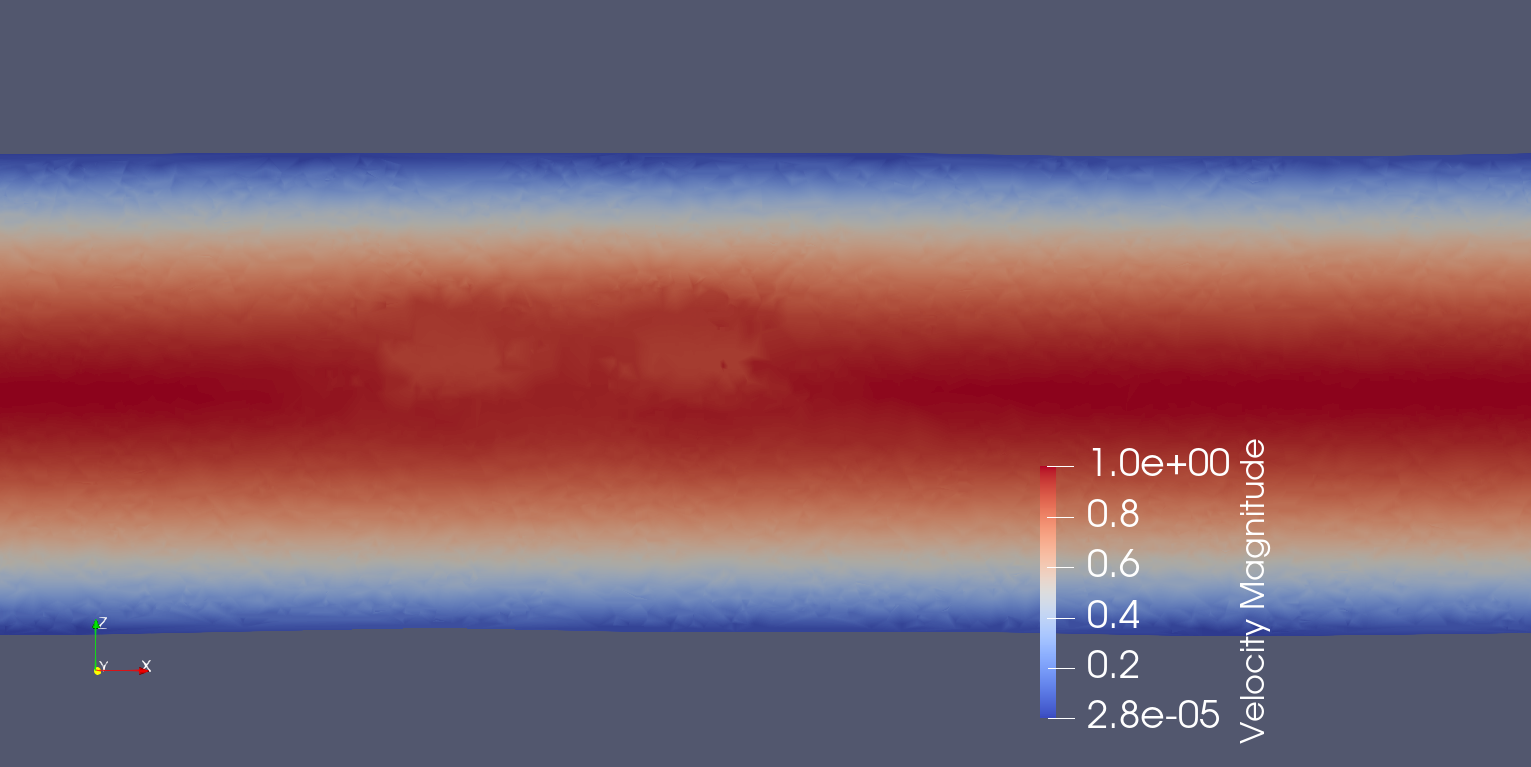}
    \endminipage       
    \caption{Variation of droplet contours with time (from top). $Cr=0.5, d_{0}=1.6$ (left column) for $ t = 1,2.2$. $Cr=0.25,d_{0}=0.5$ (right column) for $ t = 1,2.2$.}
    \label{stgrad_v_x_contour}
    \begin{picture}(1,1)
\setlength{\unitlength}{1cm}
\put(-3.2,1.0){(a)}
\put(3.6,1.0){(b)}
\end{picture}    
\end{figure}
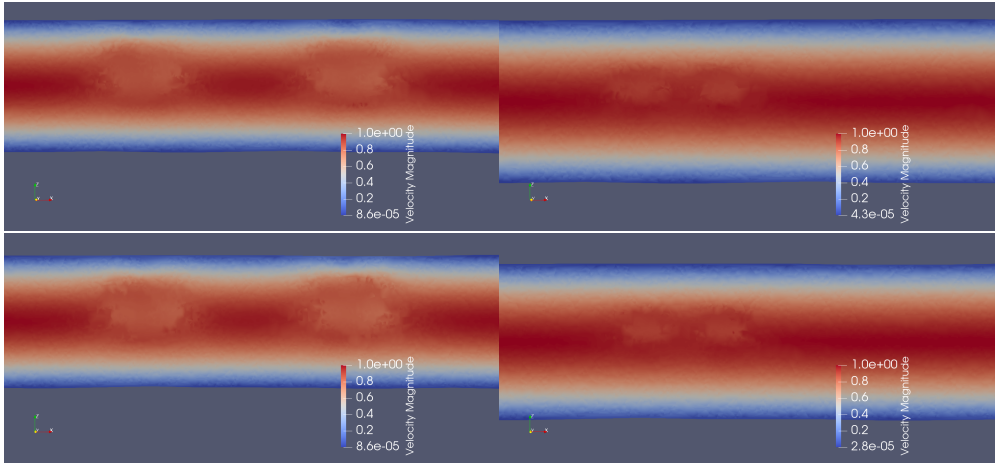
\par Overall, we have seen that the surface tension gradient on the droplet surface only plays a primary role in droplet migration in certain instances. The overall droplet migration depends not only on the surface tension gradients acting on its surface but also on the hydrodynamic effects due to the presence of the wall and the other droplets. The influence parameters determining the overall impact of the surface tension-driven forces versus the hydrodynamic forces are highly oscillatory with time for lower confinements while staying the same with time for higher confinements. Thus, at lower confinements, only one force dominates at a time, while at higher confinements, all three forces equally contribute to droplet motion.  
Thus, the confinement ratio can be seen as a significant parameter in defining the nature of the physics driving the droplet motions and is the parameter that influences the change in physics from a surface tension-driven/droplet-interaction flow to one where more droplet-interactions and wall-induced forces are involved. This is important in understanding the zones determined by non-dimensional parameters ($Cr, d_{0}$) where a more active droplet control via the thermocapillary effect can be realized.  

\subsection{Effect of the Marangoni number for different sets of conductivity ratios}
\begin{figure}
    \centering
    \minipage{0.5\textwidth}
    \includegraphics[clip,trim = 0pt 0pt 0pt 0pt,width=\linewidth]{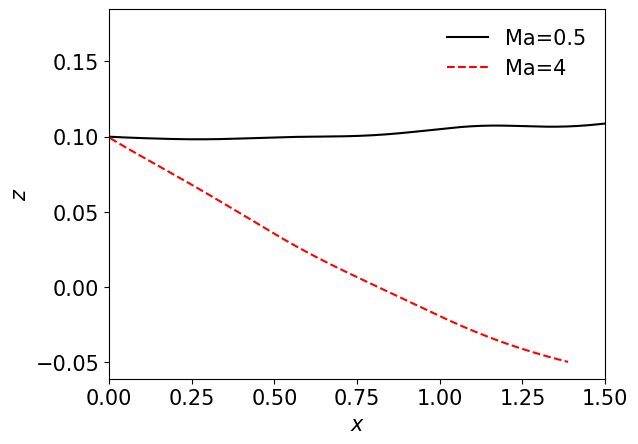}
    \endminipage\hfill
    \minipage{0.5\textwidth}
    \includegraphics[clip,trim = 0pt 0pt 0pt 0pt,width=\linewidth]{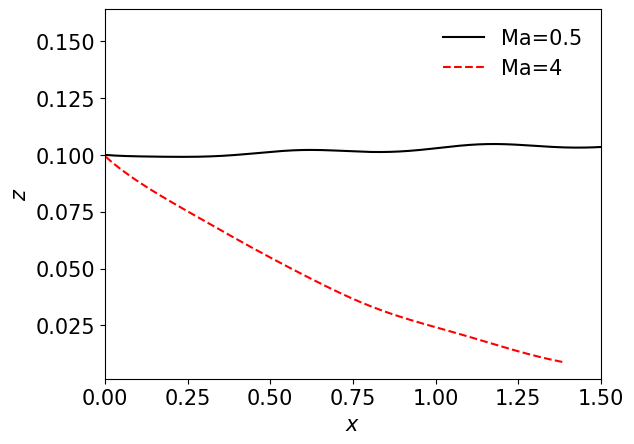}
    \endminipage
    
    \vspace*{10pt}
    \minipage{0.5\textwidth}
    \includegraphics[clip,trim = 0pt 0pt 0pt 0pt,width=\linewidth]{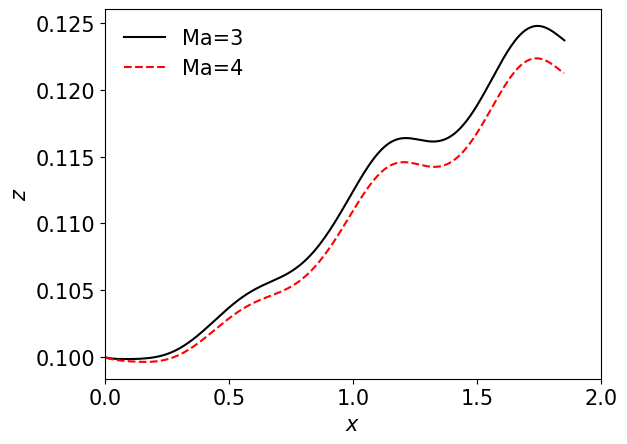}
    \endminipage\hfill
    \minipage{0.5\textwidth}
    \includegraphics[clip,trim = 0pt 0pt 0pt 0pt,width=\linewidth]{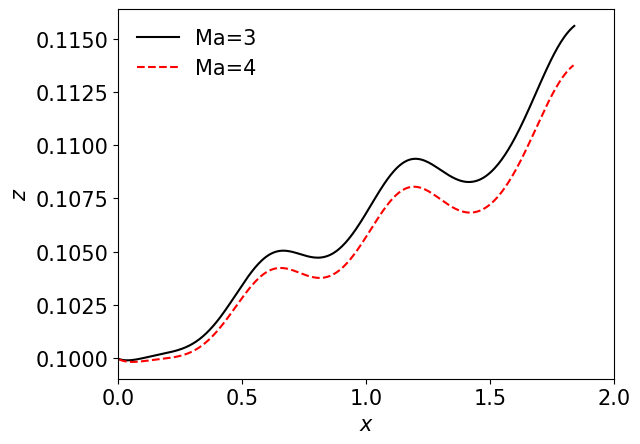}
    \endminipage
    
    \caption{Droplet centre trajectories for different Marangoni numbers ($Ma$). Top: $\delta=0.1$, bottom: $\delta=10.0$. The left column is the trailing droplet, and the right is the leading droplet.}
    \label{z_v_x_Ma}
\begin{picture}(1,1)
\setlength{\unitlength}{1cm}
\put(-3.1,6.65){(a)}
\put(3.7,6.65){(b)}
\put(-3.1,1.5){(c)}
\put(3.7,1.5){(d)}
\end{picture} 
\end{figure}
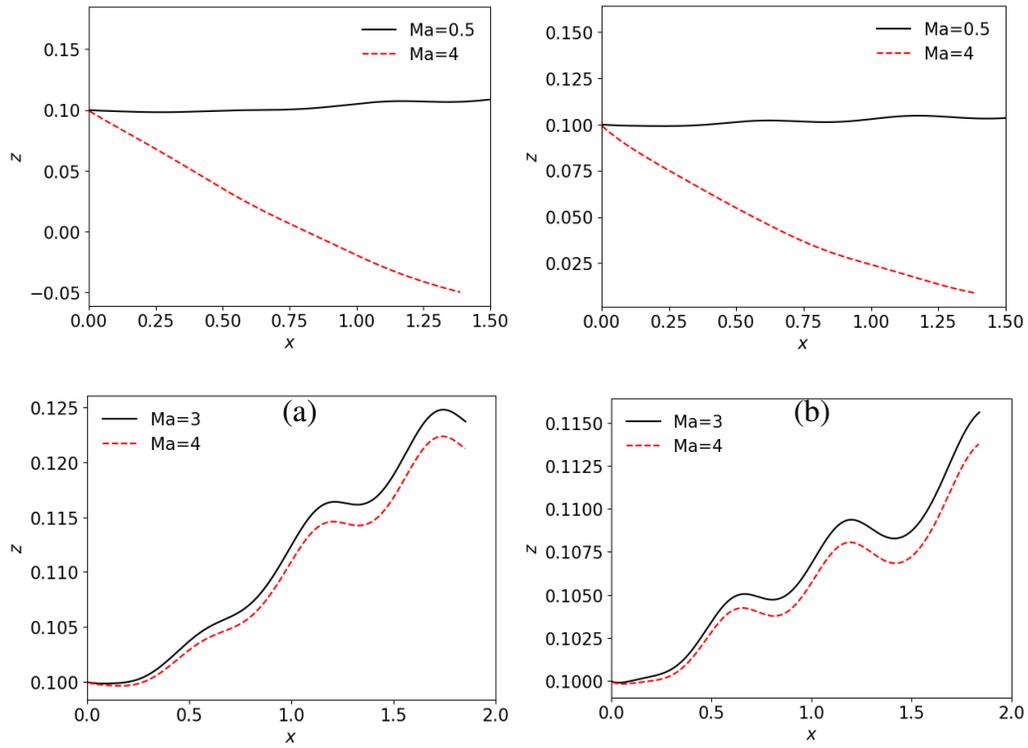

\noindent We further investigate the effect of a non-uniform thermal conductivity distribution across the droplet interface on the droplet migration. We consider two cases with thermal conductivity ratio $\delta = \overline{k}_i/\overline{k}_e = 0.1$ and $10$. We also vary the Marangoni number, keeping the confinement ratio ($Cr=0.5$) the same. The initial droplet-centre separation distance ($d_{0} = 0.55$) , the capillary number ($Ca=0.25$), and the initial $x$ and $y$ offset ($e_{l} = 0.1,p_{l} = 0$) are also kept unchanged. Changes in the Marangoni number would lead to higher thermocapillary forces. We try to see whether these increased surface tension forces become the dominant forces influencing the droplet motion at different conductivity ratios. The results for the migration trajectories for the non-uniform thermal conductivity distribution are presented in \Cref{z_v_x_Ma}

\begin{figure}
    \centering
    \minipage{0.5\textwidth}
    \includegraphics[clip,trim = 20pt 0pt 0pt 0pt,width=\linewidth]{ 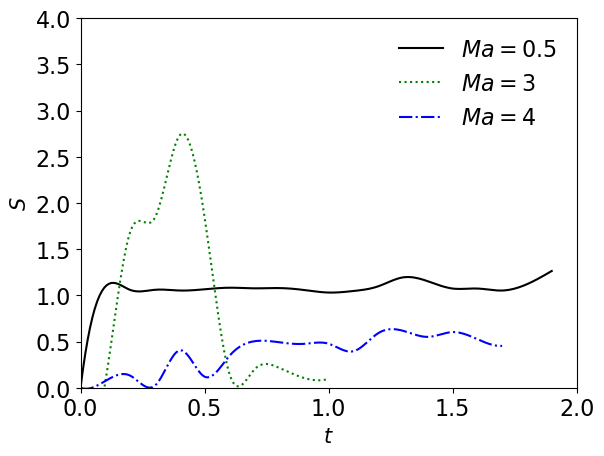}
    \endminipage\hfill
    \minipage{0.5\textwidth}
    \includegraphics[clip,trim = 25pt 0pt 0pt 0pt,width=\linewidth]{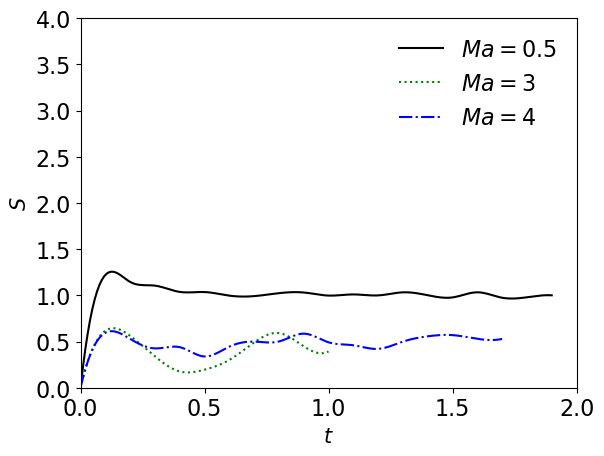}
    \endminipage

    \minipage{0.5\textwidth}
    \includegraphics[clip,trim = 20pt 0pt 0pt 0pt,width=\linewidth]{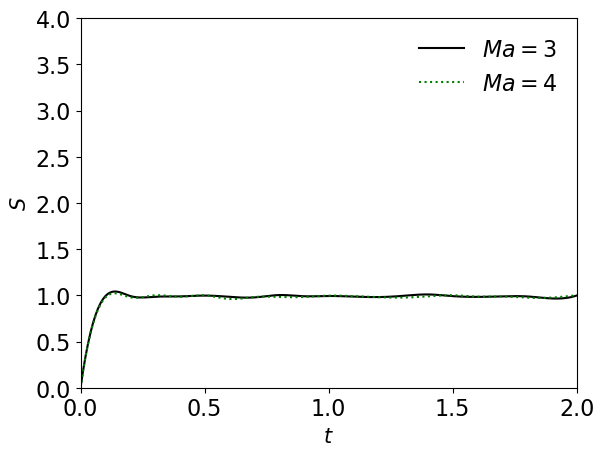}
    \endminipage\hfill
    \minipage{0.5\textwidth}
    \includegraphics[clip,trim = 25pt 0pt 0pt 0pt,width=\linewidth]{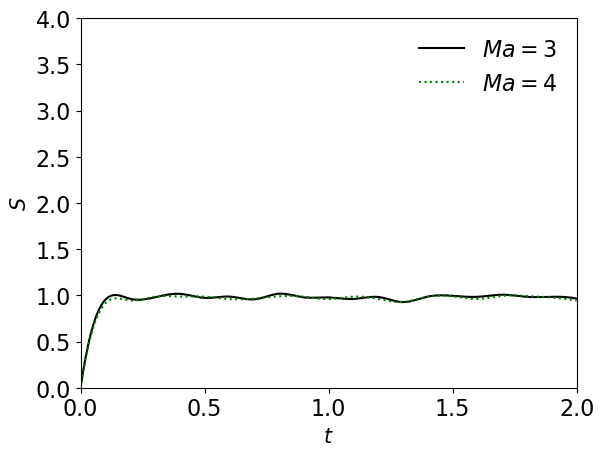}
    \endminipage
    
    \caption{Droplet influence terms (S) for different Marangoni numbers ($Ma$). Top: $\delta=0.1$, bottom: $\delta=10.0$. Left column: trailing droplet, right column: leading droplet.}
    \label{st_v_x_Ma}
    \begin{picture}(1,1)
\setlength{\unitlength}{1cm}
\put(-3.2,6.6){(a)}
\put(3.45,6.6){(b)}
\put(-3.2,1.2){(c)}
\put(3.45,1.2){(d)}
\put(-7.2,4.25){\rotatebox{90}S}
\put(-7.2,9.5){\rotatebox{90}S}
\end{picture}
\end{figure}
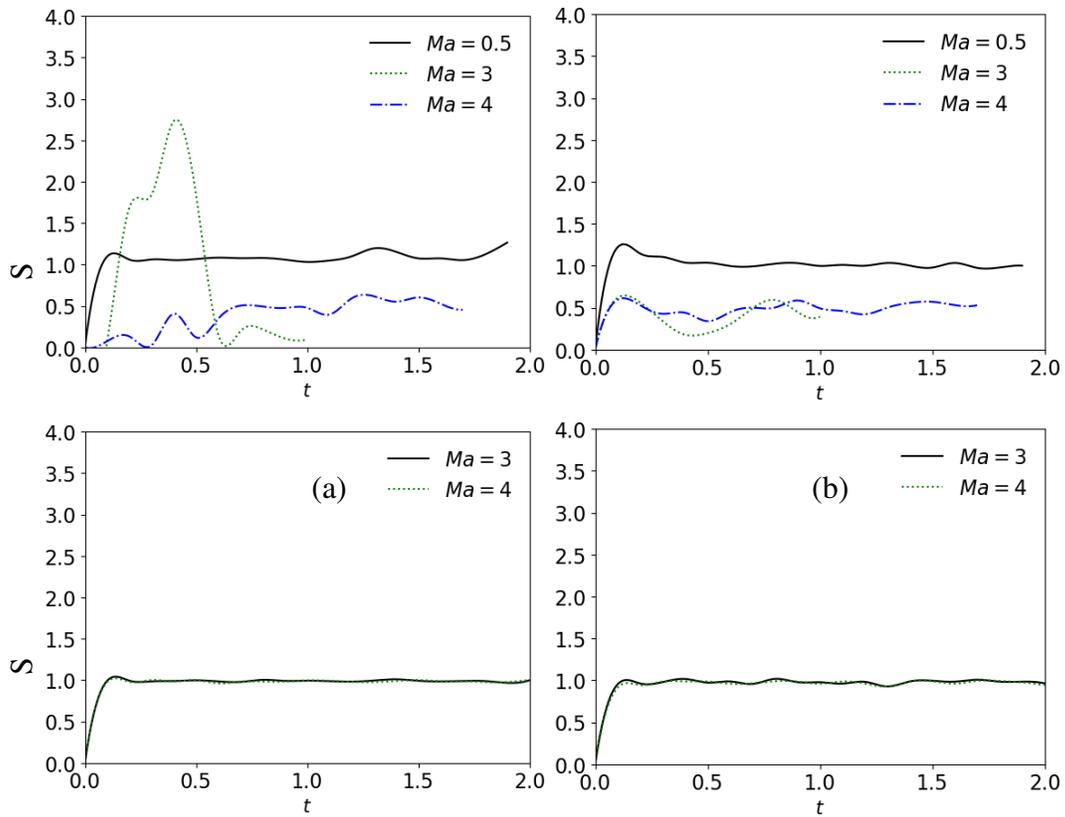

 The droplet trajectories show that with an increase in the Marangoni number, the droplet migration is skewed toward the downward direction for both values of $\delta$. However, the differences in droplet trajectory with $Ma$ are minimal for the case with a higher conductivity ratio (\Cref{z_v_x_Ma}(c),(d)). The droplet trajectories are also directed downward for lower conductivity ratios (\Cref{z_v_x_Ma}(a),(b)), while for the higher conductivity ratios (\Cref{z_v_x_Ma}(c),(d)) the motion is upward.   
 The droplet influence parameter for $Cr = 0.5,\, d_{0} = 0.55,\, g_0 = 0,\, h_0 = 0$ and different Marangoni numbers are shown in \Cref{st_v_x_Ma}. The variation in the influence parameters of the droplets for different Marangoni numbers shows that the parameters decrease with the increase in Marangoni numbers for lower conductivity ratios while remaining the same for higher ratios. For higher conductivity ratios, the influence parameters are almost one for both sets of Marangoni numbers.  

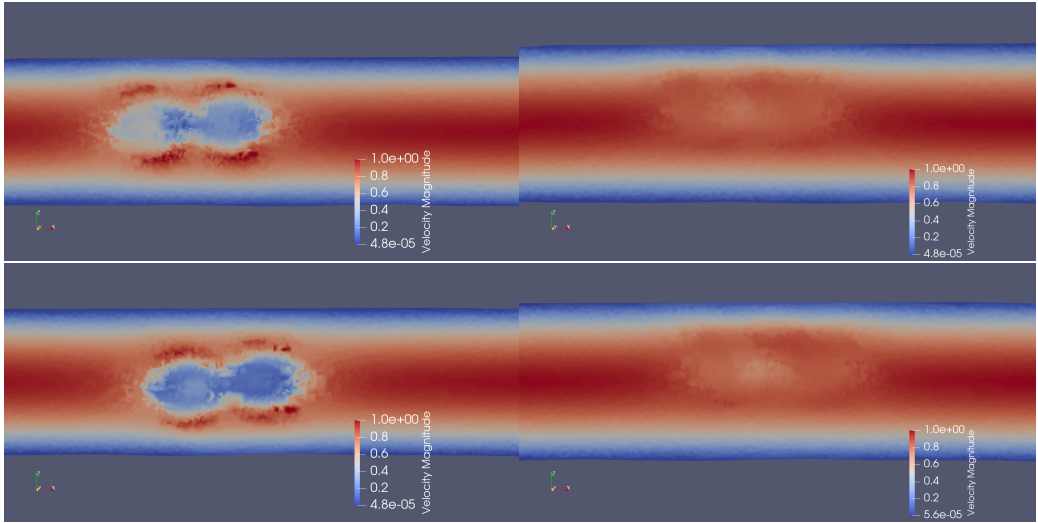
\begin{figure}
    \centering
    \minipage{0.5\textwidth}
    \includegraphics[clip,trim = 0pt 0pt 0pt 0pt,width=\linewidth]{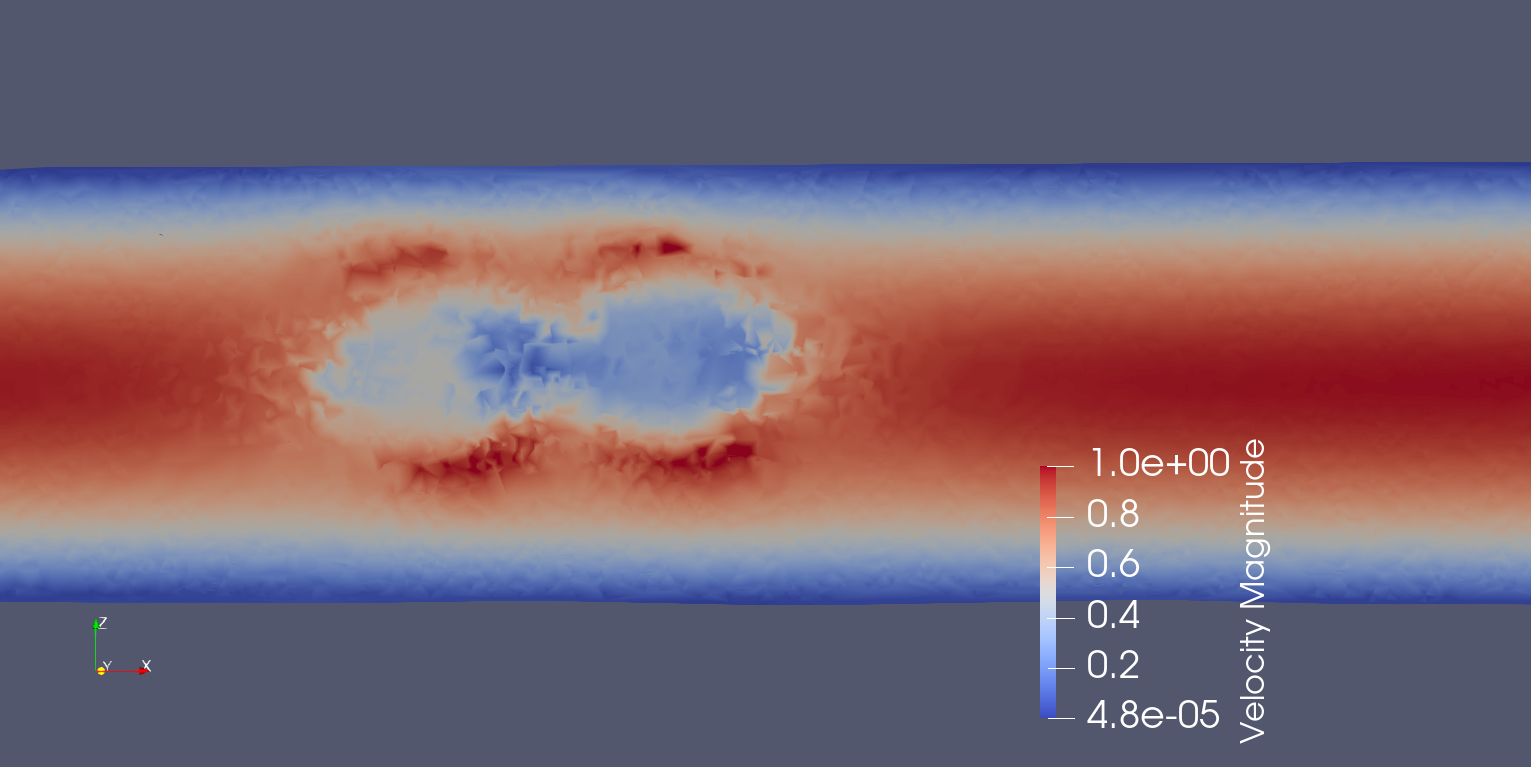}
    \endminipage\hfill
    \minipage{0.5\textwidth}
    \includegraphics[clip,trim = 0pt 0pt 0pt 0pt,width=\linewidth]
    {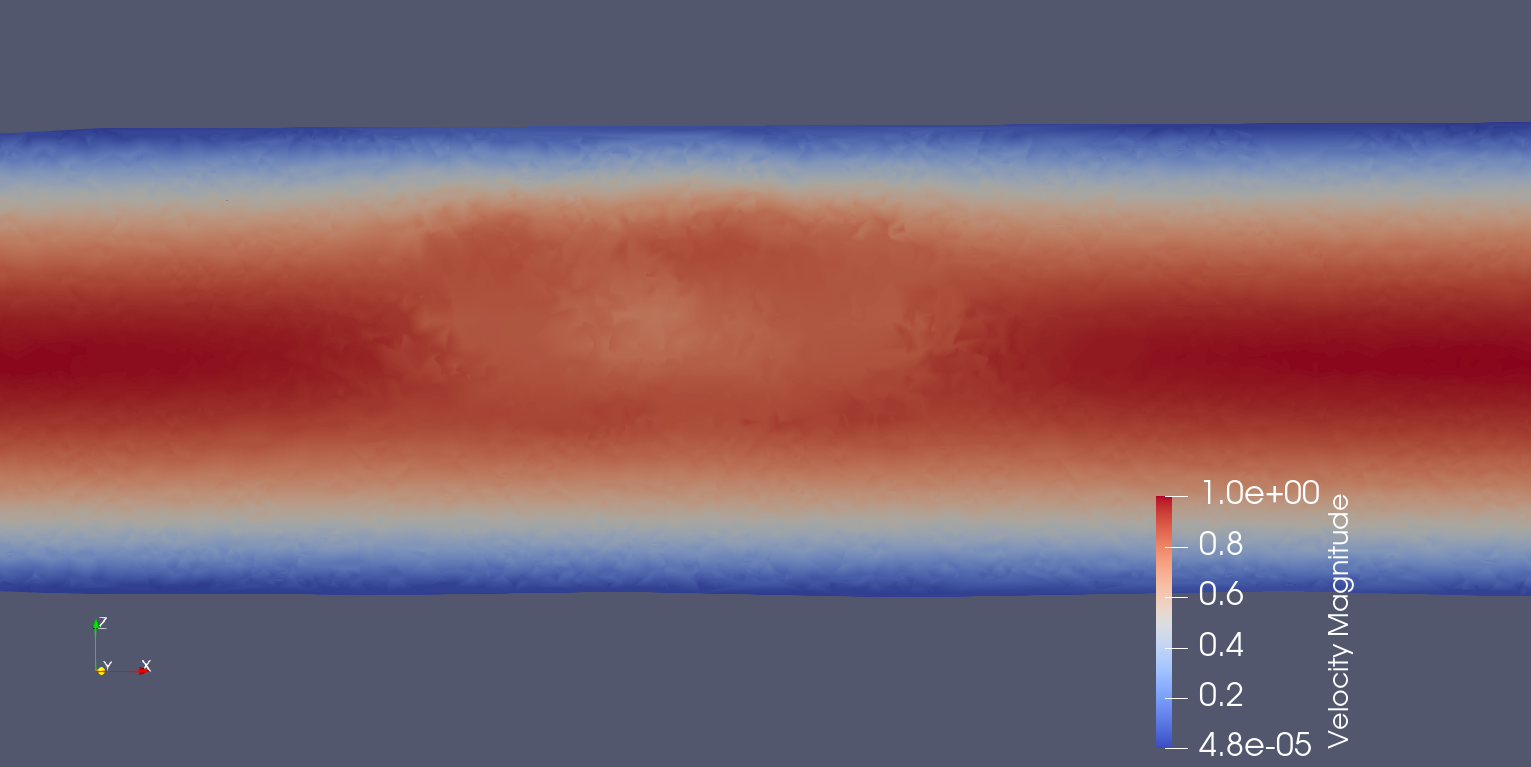}
    \endminipage

    \minipage{0.5\textwidth}
    \includegraphics[clip,trim = 0pt 0pt 0pt 0pt,width=\linewidth]{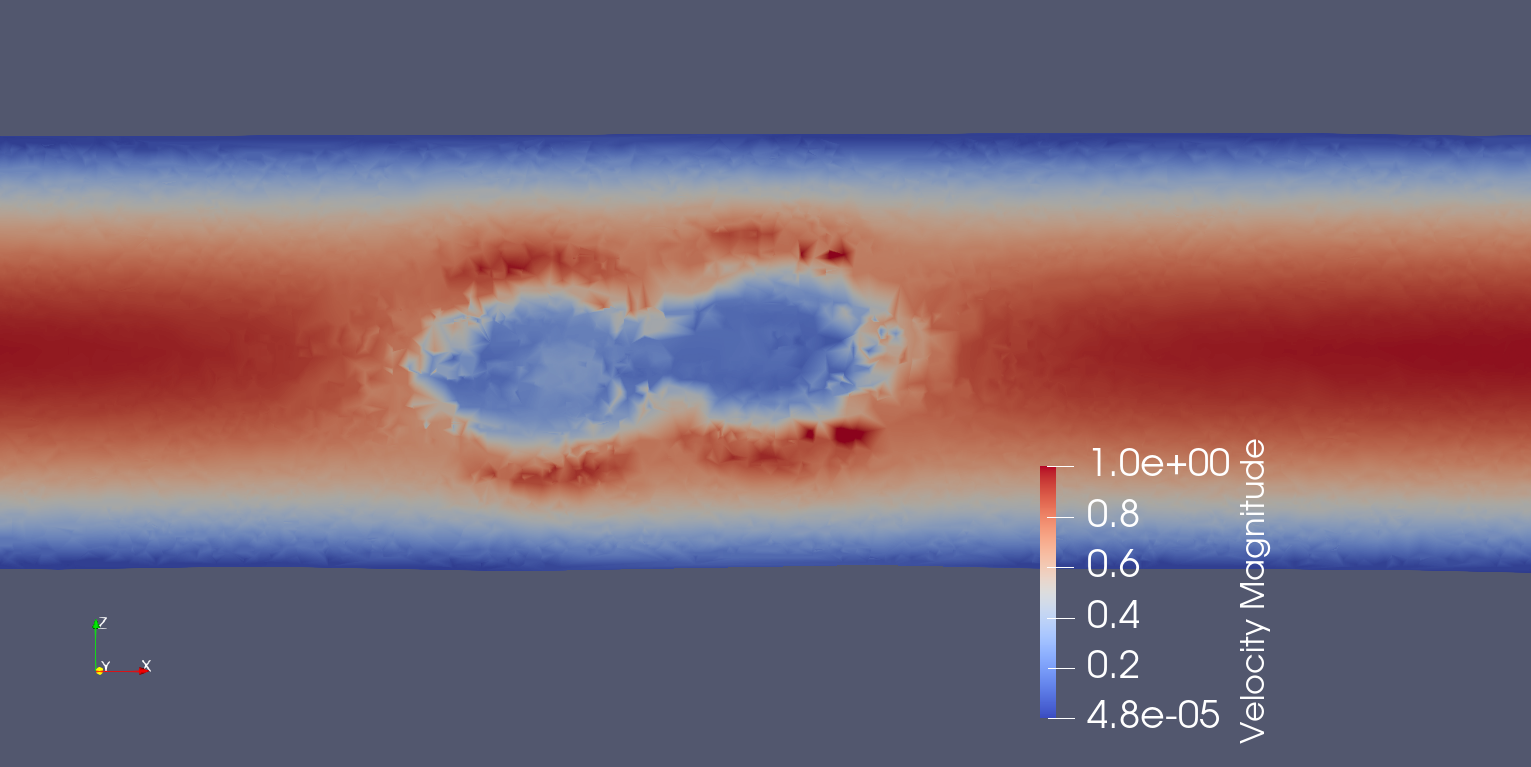}
    \endminipage\hfill
    \minipage{0.5\textwidth}
    \includegraphics[clip,trim = 0pt 0pt 0pt 0pt,width=\linewidth]{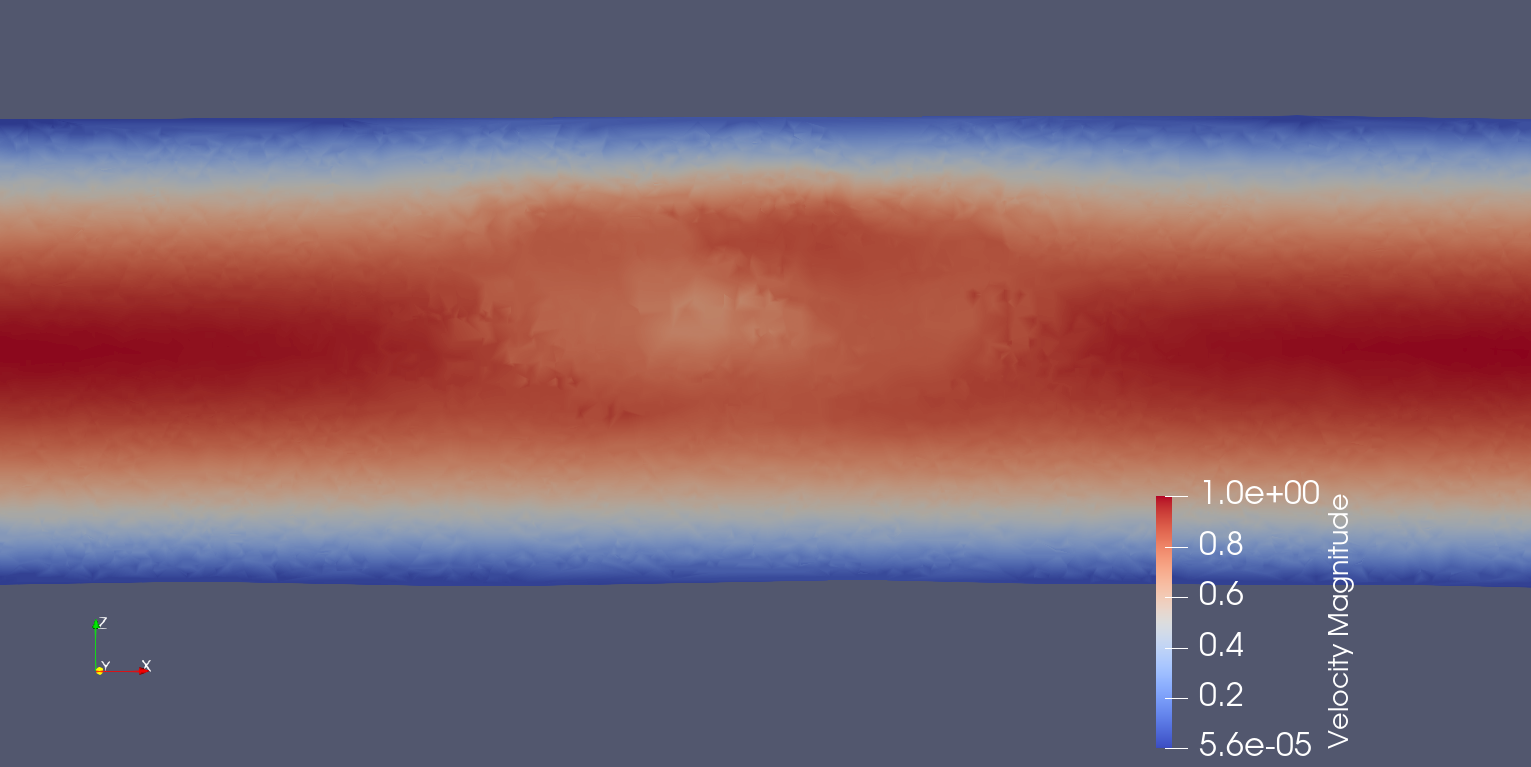}
    \endminipage
    
    \caption{Variation of droplet contours with time (from top). $Cr=0.5, d_{0}=0.55, Ma = 4$ (left column) for $ t = 0.5,1.7$. $Cr=0.5, d_{0}=0.55, Ma = 0.5$ (right column) for $ t = 0.5,1.7$.}
    \label{st_v_x_Ma_contour}
\begin{picture}(1,1)
\setlength{\unitlength}{1cm}
\put(-3.2,1.4){(a)}
\put(3.6,1.4){(b)}
\end{picture} 
\end{figure}

\noindent The conductivity ratio ($\delta = 0.1,10$) is a significant factor influencing the migration trajectory. From \Cref{fluid_bd_4}, we can see that there would be an increase in the surface tension forces with $Ma$. The downward shift in the migration trajectory with $Ma$ might be linked to this increase in Marangoni forces; however, whether these higher surface tension forces will change the physics to a surface tension-induced motion is quantified by the influence parameter ($S$) depicted in \Cref{st_v_x_Ma}. Oscillations in the influence parameters are present for lower conductivity ratios at higher $Ma$, as shown in \Cref{st_v_x_Ma}(a),(b), though they are prominent for the trailing droplet. Thus, the increase in $Ma$ at lower conductivity ratios initially causes the oscillatory behaviour seen at lower confinements for lower values of $Ma$ (\Cref{stgrad_v_x}(a),(b)). It is also seen from \Cref{st_v_x_Ma_contour} that droplet interactions are initially low but increase with time. Thus, the parameter $S$ initially oscillates from zero to high values, and these oscillations are quite high in the case of $Ma=3$ (\Cref{st_v_x_Ma}(a)). Thus, the dominance of the surface tension forces for the trailing droplet oscillates from high to low until the droplet interaction forces take over, stabilising the parameter to a steady value. In later periods of motion, the parameter($S$) reaches a stable value, though the higher Marangoni stresses ensure some degree of influence of the surface tension forces. In contrast, the influence parameter($S$) oscillates less for the leading droplet and reaches a steady state value in a shorter time (\Cref{st_v_x_Ma}(b)). 
For higher conductivity ratios (\Cref{st_v_x_Ma}(c)(d)), the overall mechanics of the droplet migration do not change much with $Ma$. It is mainly governed by the droplet interaction and wall-induced forces since the influence parameters are primarily equal to $1$, which is constant with time. The change in the conductivity ratio reduces the oscillations in the parameter $S$ for different $Ma$ in the same way as a change in the confinement ratio ($Cr$), for different initial separations ($d_0$). Thus, at higher conductivity ratios, we can assume the interaction forces and surface tension/wall forces govern the flow equally, and the increases in the $Ma$ have a minimal impact on the physics behind droplet motion. Thus, the conductivity ratio ($\delta$) is an important parameter that decides whether changes in $Ma$ affect the nature of the driving forces behind the droplet motion. 
The velocity contours for $\delta = 0.1$ are depicted in \Cref{st_v_x_Ma_contour}. These plots show strong velocity gradients at high Marangoni numbers ($Ma=4$) between the inside and outside of the droplet, unlike what we see for lower $Ma=0.5$. Strong droplet interaction is seen for both sets of $Ma$ due to the droplets being close to one another ($d_{0} = 0.55$), while for the higher $Ma$ case, they seem to increase with time, more so for the trailing droplet. The velocities inside the trailing and leading droplets are initially different, but later, they become more uniform due to the droplet interactions for $Ma=4$ (\Cref{st_v_x_Ma_contour}(a)). Initially, the droplet interactions are primarily absent, leading to surface tension forces taking precedence. These interactions take hold later, yet the surface tension effects remain, leading to steady values of the influence coefficients. The velocity inside the trailing droplet decreases with time, thus increasing the velocity gradients between itself and the external flow. For the leading droplet, the mean velocity within the droplet does not change much with time; thus, the velocity gradients between the droplet and the outside medium do not change much (\Cref{st_v_x_Ma_contour}(a)). This might be why the leading droplet's influence parameters($S$) are less oscillatory with time. 
\subsection{Effect of the Marangoni number for different sets of confinement ratios}
\noindent Now, we change the Marangoni number while also varying the confinement ratio ($Cr$) and assessing the effect of the Marangoni number on the confinement ratio and how this relates to the change in the relevant physics governing the problem. The Capillary number ($Ca = 0.25$) and the conductivity ratio  ($\delta = 10$) are kept the same while the droplet trajectories are monitored along with the surface tension gradients. The same sets of Marangoni numbers ($Ma=3,4$) and previous values of conductivity ratio ($\delta = 10$) are used for the simulations. The separation vector is given as $\mathbf{q}_0=\{d_0,0,0\}$, while the y-offset and z-offset are given as $e_l=0.1,p_l=0$. Here, $d_{0}$ and $Cr$ are changed for this study. The migration trajectories of the droplets are depicted in the following figures. 

\begin{figure}
    \centering
    \minipage{0.5\textwidth}
    \includegraphics[clip,trim = 0pt 0pt 0pt 0pt,width=\linewidth]{ 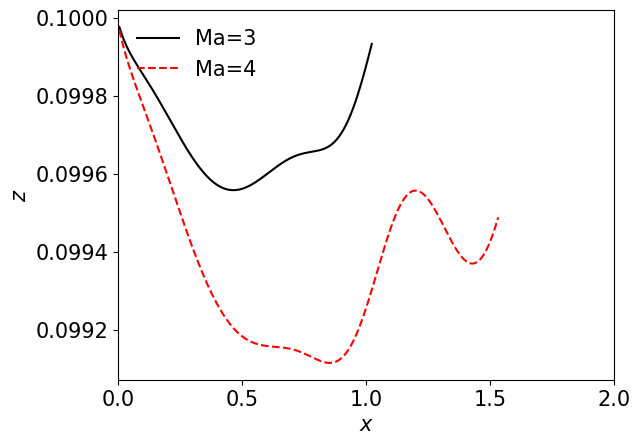}
    \endminipage\hfill
    \minipage{0.5\textwidth}
    \includegraphics[clip,trim = 0pt 0pt 0pt 0pt,width=\linewidth]{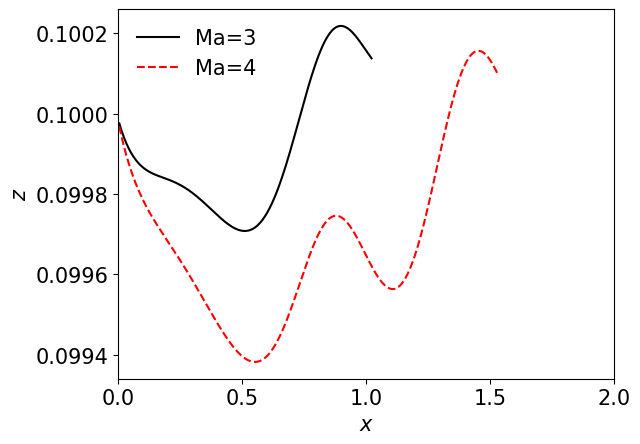}
    \endminipage
    \vspace*{10pt}
    \minipage{0.5\textwidth}
    \includegraphics[clip,trim = 0pt 0pt 0pt 0pt,width=\linewidth]{Ma_z_x.png}
    \endminipage\hfill
    \minipage{0.5\textwidth}
    \includegraphics[clip,trim = 0pt 0pt 0pt 0pt,width=\linewidth]{Ma_z_x_leading.png}
    \endminipage
    \vspace*{10pt}
    \caption{Droplet centre trajectories for different Marangoni numbers. Top: ($Cr=0.25$, $d_{0}=0.3$) Bottom: ($Cr=0.5$, $d_{0}=0.55$) Left: trailing droplet, right: leading droplet.}
    \label{st_v_x_Cr}
\begin{picture}(1,1)
\setlength{\unitlength}{1cm}
\put(-3.1,6.65){(a)}
\put(3.7,6.65){(b)}
\put(-3.1,1.5){(c)}
\put(3.7,1.5){(d)}
\end{picture} 
\end{figure}
\begin{figure}
    \centering
    \minipage{0.5\textwidth}
    \includegraphics[clip,trim = 20pt 0pt 0pt 0pt,width=\linewidth]{ 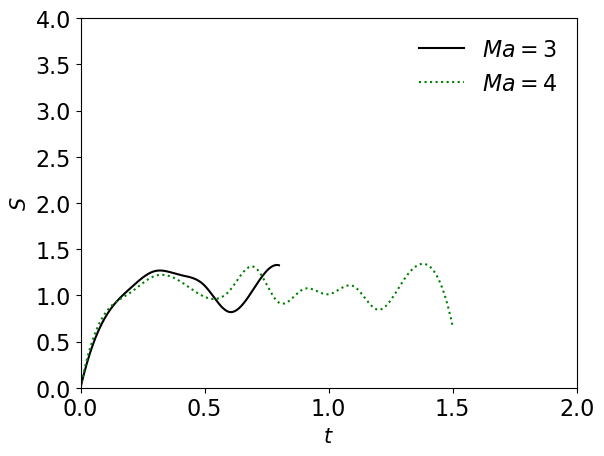}
    \endminipage
    \minipage{0.5\textwidth}
    \includegraphics[clip,trim = 25pt 0pt 0pt 0pt,width=\linewidth]{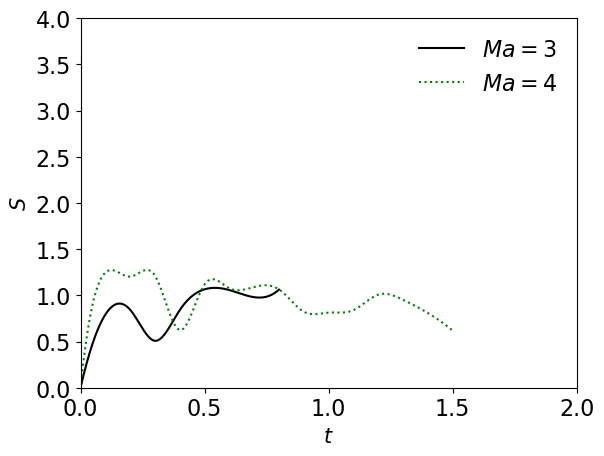}
    \endminipage

    \minipage{0.5\textwidth}
    \includegraphics[clip,trim = 20pt 0pt 0pt 0pt,width=\linewidth]{srftn_trailing_Ma_3_42.png}
    \endminipage\hfill
    \minipage{0.5\textwidth}
    \includegraphics[clip,trim = 25pt 0pt 0pt 0pt,width=\linewidth]{srftn_leading_Ma_3_42.png}
    \endminipage
    
    \caption{Droplet influence coefficients for different Marangoni numbers. Top: ($Cr = 0.25,d_{0} = 0.3$) Bottom: ($Cr = 0.5,d_{0} = 0.55$) Left: Trailing droplet right: Leading droplet.}
    \label{st_v_x_Ma_confine}
        \begin{picture}(1,1)
\setlength{\unitlength}{1cm}
\put(-3.1,7.11){(a)}
\put(3.6,7.10){(b)}
\put(-3.2,1.5){(c)}
\put(3.6,1.5){(d)}
\put(-7.2,5){\rotatebox{90}S}
\put(-7.2,10.5){\rotatebox{90}S}
\end{picture}
\end{figure}

The migration trajectories for the droplets are shown in \Cref{st_v_x_Cr}. For the lower confined case (\Cref{st_v_x_Cr}(a),(b)), it can be seen that the droplet moves downward and then goes upward, while for the higher confined case (\Cref{st_v_x_Cr}(c), (d)), the motion is entirely upward. Also, it can be seen that for the lower confined case, the droplet migration changes considerably with $Ma$, unlike what we see at higher confinements. 
The droplet migration is a result of the interplay between the pressure gradients generated by the surface tension gradients on their surface and the hydrodynamic lift forces resulting from their interactions with one another and the walls. As we have seen previously, in some cases, the former plays a major role in influencing the droplet trajectory, while the latter plays a prominent role in other cases. Thus, we have the influence coefficients depicted in \Cref{st_v_x_Ma_confine}, and unlike what we see for lower conductivity ratios ($\delta = 0.1$), the values are much higher and close to $1$. From \Cref{st_v_x_Ma_confine}(a),(b), we see that in the initial range of motion for ($Cr=0.125$), the influence coefficients for both Marangoni numbers remain the same for the trailing droplet but increase with the Marangoni number for the leading droplet in the first few seconds of motion. This shows that the variation in the surface tension gradient or the thermocapillary forces influences the trajectory of the leading droplet differently compared to the trailing droplet. Thus, we can see that at higher conductivity ratios, the increase in $Ma$ has a slight effect on the physics behind the droplet motion at lower confinements. Thus, the $Ma$ can change the driving forces behind droplet motion more so at lower conductivities or at lower confinements.

\noindent Here from \Cref{st_v_x_Ma_confine}, we can see that for higher confinements ($Cr=0.5$), the motion is almost entirely independent of the surface tension gradient evolution. The influence coefficients are close to $1$, meaning the surface tension forces scarcely affect the overall motion. The motion is due to more complex physics depending on the droplet deformation and the wall and droplet interactions that are more magnified at such high Marangoni numbers. Thus, at higher Marangoni numbers, higher conductivity ratios, and higher confinements, the problem's physics is more linked to the complex interplay of the hydrodynamic fields and the resulting droplet deformation rather than the thermocapillary forces. For lower confinements ($Cr = 0.25$), the Marangoni number has some effect on the flow physics, at least for the leading droplet. However, the impact of surface tension gradients on the droplet surface is still much lower compared to the case with a lower conductivity ratio, as discussed in the previous section. This shows that along with the Marangoni number, the conductivity ratio plays a role in deciding whether the Marangoni number can alter the dynamics of droplet motion.
\subsection{Effect of displacement in the y-direction}
Now, we look at the effects of droplet separation in the y-direction. Initial parameters like $Ca = 0.25,\delta = 0.1, Ma = 0.5, Cr = 0.25$ are the same as previous simulations. The separation in the y-direction is introduced. The separation vector is given as $\mathbf{q}_0=\{d_0,g_0,0\}$. Here, two cases are presented for $d_0 = 0.5,\,g_0 = 0.6,\,e_l=0.1,\,p_1=0.3,\,p_2=-0.3$ and $d_0 = 0.5,\,g_0 = 0.84,\,e_l=0.1,\,p_1=-0.5,\,p_2=0.34$. The droplet trajectories are presented in \Cref{st_v_x_Ma_confine_y_d1} and \Cref{st_v_x_Ma_confine_y_d2}. 
\begin{figure}
    \centering
    \minipage{0.5\textwidth}
    \includegraphics[clip,trim = 0pt 0pt 0pt 0pt,width=\linewidth]{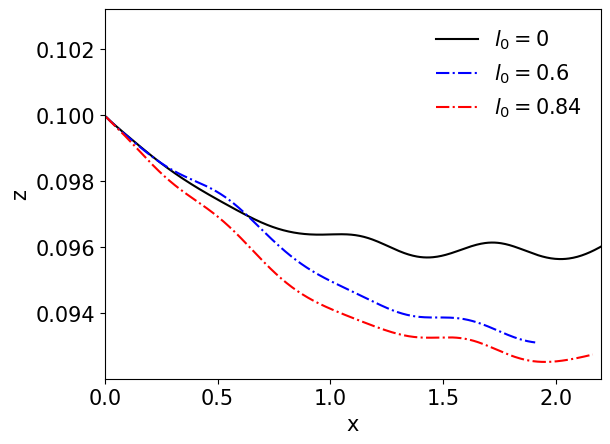}
    \endminipage
    \minipage{0.5\textwidth}
    \includegraphics[clip,trim = 0pt 0pt 0pt 0pt,width=\linewidth]{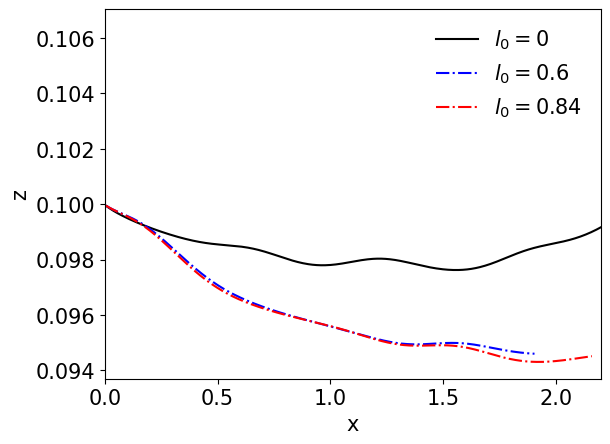}
    \endminipage

    \caption{Droplet center trajectories for different y-separations. Left: Trailing droplet right: Leading droplet.}
    \label{st_v_x_Ma_confine_y_d1}
        \begin{picture}(1,1)
\setlength{\unitlength}{1cm}
\put(-3.1,1.11){(a)}
\put(3.6,1.10){(b)}
\end{picture}
\end{figure}
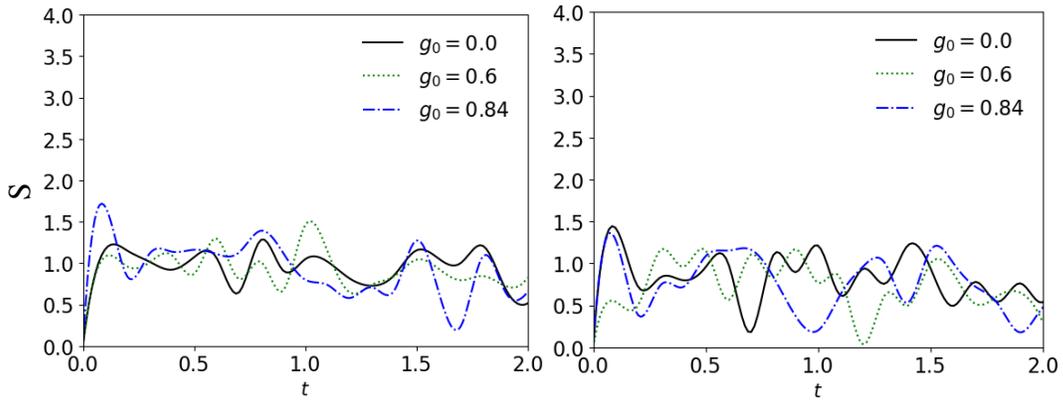
\begin{figure}
    \centering
    \minipage{0.5\textwidth}
    \includegraphics[clip,trim = 20pt 0pt 0pt 0pt,width=\linewidth]{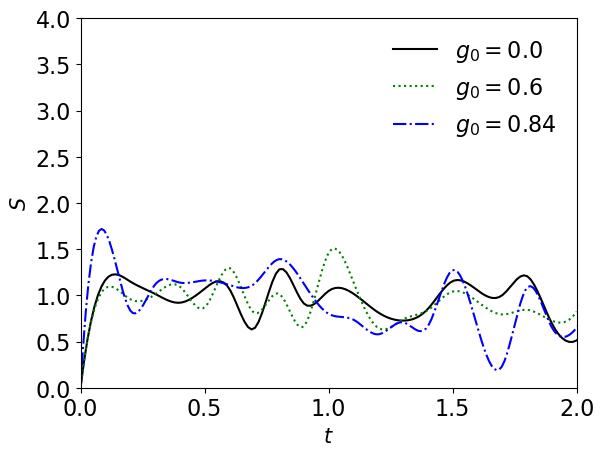}
    \endminipage\hfill
    \minipage{0.5\textwidth}
    \includegraphics[clip,trim = 25pt 0pt 0pt 0pt,width=\linewidth]{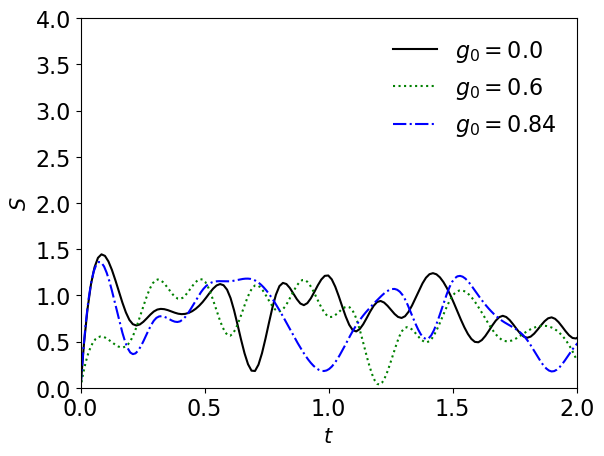}
    \endminipage    
    \caption{Droplet influence coefficients for different y-separations. Left: Trailing Droplet right: Leading Droplet.}
    \label{st_v_x_Ma_confine_y_d2}
        \begin{picture}(1,1)
\setlength{\unitlength}{1cm}
\put(-3.1,1.11){(a)}
\put(3.6,1.10){(b)}
\put(-7.2,4.5){\rotatebox{90}S}
\end{picture}
\end{figure}
The trajectories of droplet motion are given in \Cref{st_v_x_Ma_confine_y_d1}. The droplet motion is directed downward with the increase in $y$-separation. For the trailing droplet, there seems to be some difference in the trajectories for the two y-separation values considered. For the leading droplet, there is very little variation in the trajectories as the $y$- separation increases. The influence coefficients are also plotted in \Cref{st_v_x_Ma_confine_y_d2}. As with the case with no $y$-separation ($g_0 = 0.0$), the influence coefficients are oscillatory with time; however, the change in the mean value is marginal. For the trailing droplet, the oscillations in the case of $g_0=0.6,0.84$ are higher than $g_0=0$. In the case of the leading droplet, the oscillations are similar for all values of $g_0$. For the leading droplet, there are instances when the influence coefficients reach a very low value close to zero for all three $y$-separations considered. The separation between the droplets in the $y$-direction is seen to have a very limited effect on the role of the droplet interaction forces.

\section{Conclusions}
We investigated the behaviour of multiple droplets in a confined domain, focusing on how specific physical parameters might uniquely influence their strongly coupled motion. The analysis was conducted in regimes characterized by infinitesimally small Reynolds number and Peclet number and for small values of the Capillary number, using the boundary element method. The migration behaviour of the droplets was shown to depend on several factors, including the
confinement ratio ($Cr$), thermal Marangoni number ($Ma$), conductivity ratio between the continuous fluid and the droplets ($\delta$), and the initial separation distance ($d_0$) between droplets.

Our findings highlighted the importance of an \emph{influence parameter} (S), delineating the confluence of the surface tension forces, droplet interaction forces stemming from the effect of the surface temperature gradient on the other droplet, and the wall-induced forces due to the presence of the confining boundaries, in determining the governing
physics of droplet dynamics. Notably, at higher conductivity ratios, changes in the Marangoni number ($Ma$) were shown to have minimal effects on the overall droplet motion. The leading droplet experienced greater propulsion due to surface
tension-induced forces, while the trailing droplet moved within the thermal wake of the leading one. The dynamics were illustrated through a depiction of the flow field, showcasing constant velocity gradients in the leading droplet as compared to the trailing one. The thermocapillary-induced surface tension force is observed to be the key factor for
droplet migration. By altering the local velocity field and hence the wall-induced lift, this could also have an implicit handle on the hydrodynamic forces, which would selectively favour or oppose the interfacial forces depending on the parametric regime. The cumulative influence was thus illustrated to be influenced by the range of the following sets of parameters in combination: ($Cr$, $d_0$, $Ma$, $\delta$).

Nevertheless, within the scope of the physics addressed in this work, the confinement ratio, Marangoni number, and conductivity ratio were shown
to be the key dimensionless parameters influencing droplet migration. At lower conductivity and confinement ratios, thermocapillary forces
driven by surface tension gradients, and the droplet-interaction forces were shown to dominate the motion in an alternate manner. Conversely, at higher conductivity and confinement ratios, the wall-induced lift forces or droplet interaction forces were deciphered to become the dominant drivers. The Marangoni number ($Ma$) was revealed to play a more significant role at lower conductivity ratios and lower confinements, with higher values of $Ma$ enhancing surface tension-driven motion in such cases. While the results illustrated in this work were aimed at capturing the trends in droplet motion based on a limited range of parameter values deciphering the specific physics of interest, broader parametric sweeps incorporating a wider spectrum of non-dimensional parameters may indeed unveil other migration patterns, which may have exclusive implications depending on the specific application on hand.

The conceptual insights presented here could be translated to the practical realm by leveraging phenomenal recent advancements in miniaturization and space technology, where precise control of droplet migration is crucial for numerous applications in near-weightless environments. For instance, the use of thermocapillary forces to remove unwanted liquid drops from a continuous phase could significantly enhance materials processing in outer space, mitigating defects commonly caused by gravity-induced fluid phase segregation. With the progress of microfluidics and miniaturization, thermocapillary phenomena are garnering attention for other emerging terrestrial applications as well.

For example, micro heat pipes used in the thermal management of electronic devices benefit from these phenomena due to their high surface-area-to-volume ratios. Thermocapillary control of multiple-droplet systems might form the basis of their cooling system design, albeit with judicious process control. This is because, in practice, uncontrolled droplet accumulation driven by thermocapillary motion may reduce the efficiency of heat exchange between hot and cool interfaces. Thus, achieving precise control over thermocapillary migration - whether to accelerate or
decelerate droplet motion - is imperative. The present study may therefore serve as an important precursor toward developing such control
mechanisms, paving the way for advancements in both space and terrestrial applications for the future.
\section{}\label{appA}
\subsection{Validation}
\begin{figure}
\centering
\begin{subfigure}{0.5\textwidth}
\centering
\includegraphics[width=0.9\textwidth]{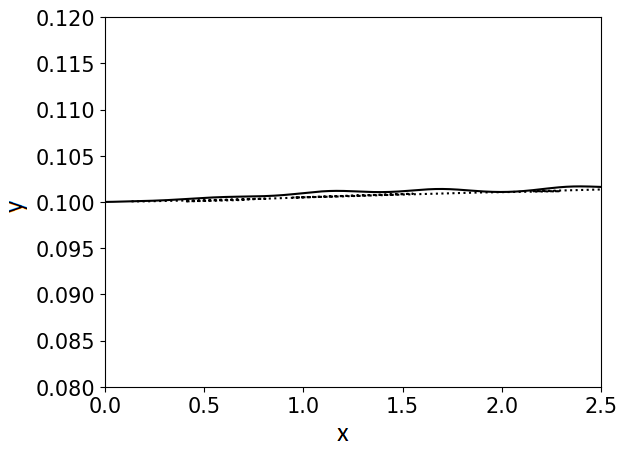}  
\end{subfigure}%
\begin{subfigure}{0.5\textwidth}
\centering
\includegraphics[width=0.9\textwidth]{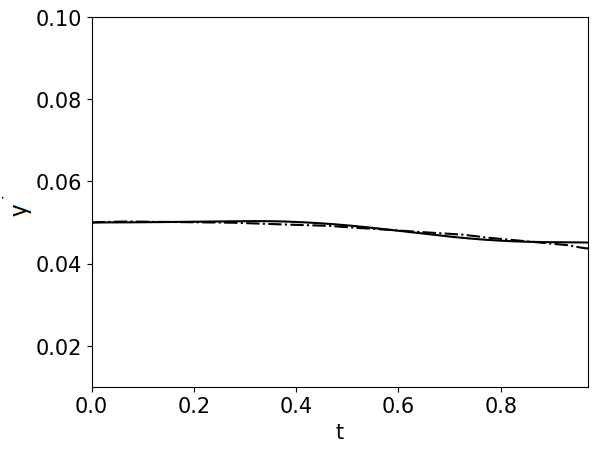}  
\end{subfigure}
\centering
\caption{Droplet centre trajectory of the single droplet in the absence of temperature gradient validated with \citet{Mortaza2002} Left:$Cr = 0.25$ Right:$Cr = 0.75$}
\label{fig:figure1}
\end{figure}

\begin{figure}
     \centering
     \minipage{0.5\textwidth}
     \centering
    \includegraphics[clip,trim = 0pt 0pt 0pt 10pt,width=0.9\textwidth]{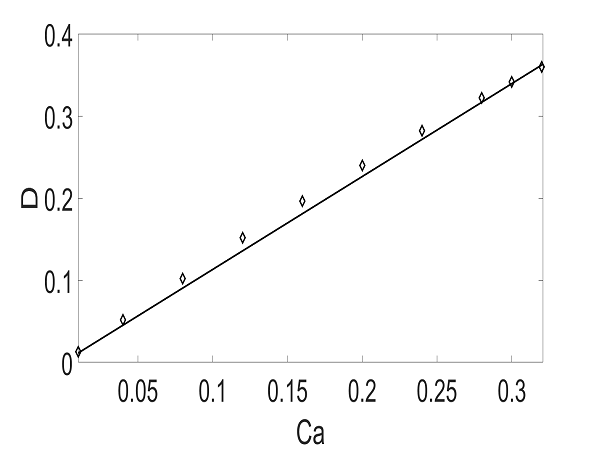}      
     \endminipage%
     \minipage{0.5\textwidth}
     \centering
    \includegraphics[width=0.9\textwidth]{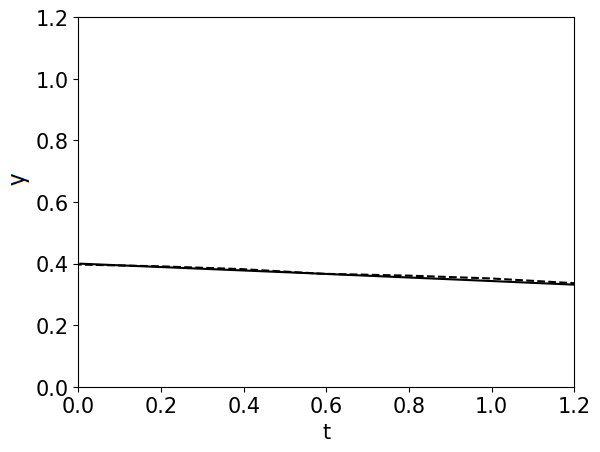}      
     \endminipage%
     \centering
     \caption{Left: Droplet deformation with capillary number Right: Droplet centre trajectory validated with \citet{Santra2023}}
     \label{fig:figure7}
\end{figure}

To assess the method's accuracy, we considered the migration of a single droplet in the presence of a Posieulle flow bounded by no-slip walls in an isothermal domain, as discussed in \citet{Mortaza2002}. We assume low Reynolds numbers and Capillary numbers, and the droplet viscosity is considered to be the same as that of the bulk fluid. In the limit of low capillary and Reynolds numbers, the deformation of the droplet is assumed to be negligible while the flow is within the Stokes flow regime. The capillary number is defined as \begin{equation*}
    Ca=\frac{U_{0}\mu_{o}}{\sigma}
\end{equation*}
The non-dimensional droplet radius ($a$) is taken to be $0.125H$. The capillary number ($Ca$) has been taken to be $0.25$. 
The droplet is positioned at a distance above the centreline of the channel, which is taken to be $0.1$. In the first figure, we see the migration of the drop in the absence of a thermal gradient. The locus of the droplet centroid is compared with those of \cite{Mortaza2002} as depicted in \Cref{fig:figure1}. The overall error is of the order of around less than $0.95\%$.  
\\Next, we tested whether our model can reproduce the effects of confinement on a single droplet by changing the confinement ratio ($Cr$) to $0.75$, while not changing any of the other non-dimensional parameters ($Ca,a$). This completes the validation of our single droplet model with previous similar studies employing different computational techniques.
For the case of multiple droplets, we considered a pair of droplets moving in uniform shear flow to validate it with a study by Rallison(1981). Droplet deformation is plotted against the capillary number, and we see reasonable agreement with our results and those of Rallison(1981). Next, the coupled heat transfer-fluid flow problem is solved in the regime of low Reynolds, Peclet, and Capillary numbers to validate the model with the results of \citet{Santra2023}. The latter uses different computational methods and solves the problem for a transverse temperature field and different values of Marangoni and Peclet numbers. The results are shown in \Cref{fig:figure7}, and in the realm of low Reynolds and Peclet numbers, both results are almost coincident with the calculated error being less than $0.9\%$.

As can be seen, our results are in good agreement with several previous computational studies employing different computational techniques. 

\bibliographystyle{jfm}
\bibliography{main_file}

\end{document}